# CONSISTENCY OF SUPPORT VECTOR MACHINES FOR FORECASTING THE EVOLUTION OF AN UNKNOWN ERGODIC DYNAMICAL SYSTEM FROM OBSERVATIONS WITH UNKNOWN NOISE

By Ingo Steinwart and Marian Anghel

*Los Alamos National Laboratory*

We consider the problem of forecasting the next (observable) state of an unknown ergodic dynamical system from a noisy observation of the present state. Our main result shows, for example, that support vector machines (SVMs) using Gaussian RBF kernels can learn the best forecaster from a sequence of noisy observations if (a) the unknown observational noise process is bounded and has a summable $\alpha$-mixing rate and (b) the unknown ergodic dynamical system is defined by a Lipschitz continuous function on some compact subset of $\mathbb{R}^d$ and has a summable decay of correlations for Lipschitz continuous functions. In order to prove this result we first establish a general consistency result for SVMs and all stochastic processes that satisfy a mixing notion that is substantially weaker than $\alpha$-mixing.

Let us assume that we have an ergodic dynamical system described by the sequence $(F^n)_{n \geq 0}$ of iterates of an (essentially) unknown map $F : M \to M$, where $M \subset \mathbb{R}^d$ is compact and the corresponding ergodic measure $\mu$ is assumed to be unique. Furthermore, assume that all observations $\tilde{x}$ of this dynamical system are corrupted by some stationary, $\mathbb{R}^d$-valued, additive noise process $\mathcal{E} = (\varepsilon_n)_{n \geq 0}$ whose distribution $\nu$ we assume to be independent of the state, but otherwise *unknown*, too. In other words *all possible* observations of the system at time $n \geq 0$ are of the form

$$\tilde{x}_n = F^n(x_0) + \varepsilon_n, \tag{1}$$

where $x_0$ is a true but unknown state at time 0. Now, given an observation of the system at some arbitrary time, our goal is to forecast the next *observable*

Received April 2007; revised October 2007.

state (we will see later that under some circumstances this is equivalent to forecasting the next *true* state), that is, given $x + \varepsilon$ we want to forecast $F(x) + \varepsilon'$, where $\varepsilon$ and $\varepsilon'$ are the observational errors for $x$ and its successor $F(x)$. Of course, if we know neither $F$ nor $\nu$, then this task is impossible, and hence we assume that we have a finite sequence $T = (\tilde{x}_0, \ldots, \tilde{x}_{n-1})$ of noisy observations from a trajectory of the dynamical system, that is, all $\tilde{x}_i$, $i = 0, \ldots, n-1$, are given by (1) for a conjoint initial state $x_0$. Now, informally speaking, our goal is to use $T$ to build a forecaster $f : \mathbb{R}^d \to \mathbb{R}^d$ whose average forecasting performance on future noisy observations is as small as possible. In order to render this goal more precisely we need a loss function $L : \mathbb{R}^d \to [0, \infty)$ such that

$$L(F(x) + \varepsilon' - f(x + \varepsilon))$$

gives a value for the discrepancy between the forecast $f(x + \varepsilon)$ and the observed next state $F(x) + \varepsilon'$. In the following, we always assume implicitly that small values of $L(F(x) + \varepsilon' - f(x + \varepsilon))$ correspond to small values of $\|F(x) + \varepsilon' - f(x + \varepsilon)\|_2$, where $\| \cdot \|_2$ denotes the Euclidean distance in $\mathbb{R}^d$. Now, by the stationarity of $\mathcal{E}$, the average forecasting performance is given by the $L$-risk

$$(2) \qquad \mathcal{R}_{L,P}(f) := \iint L(F(x) + \varepsilon_1 - f(x + \varepsilon_0)) \, \nu(d\varepsilon) \mu(dx),$$

where $\varepsilon = (\varepsilon_i)_{i \geq 0}$ and $P := \nu \otimes \mu$. Obviously, the smaller the risk the better the forecaster is, and hence we ideally would like to have a forecaster $f_L^* : \mathbb{R}^d \to \mathbb{R}^d$ that attains the minimal $L$-risk

$$(3) \qquad \mathcal{R}_{L,P}^* := \inf\{\mathcal{R}_{L,P}(f) | f : \mathbb{R}^d \to \mathbb{R}^d \text{ measurable}\}.$$

Now assume that we have a method $\mathcal{L}$ that assigns to every training set $T$ a forecaster $f_T$. Then the method $\mathcal{L}$ achieves our goal asymptotically, if it is *consistent* in the sense of

$$(4) \qquad \lim_{n \to \infty} \mathcal{R}_{L,P}(f_T) = \mathcal{R}_{L,P}^*,$$

where the limit is in probability $P$.

To the best of our knowledge, the forecasting problem described by (1)–(4) has not been considered in the literature, and even the observational noise model itself has only been considered sporadically, though it clearly "captures important features of many experimental situations" [27]. Moreover, most of the existing work on the observational noise model deals with the question of denoising [17, 23, 24, 25, 26, 27, 35]. In particular, [25, 26, 27] provide both positive and negative results on the existence of consistent denoising procedures.

In [32] a related forecasting goal is considered for the least squares loss and stochastic processes of the form $Z_{i+1} := F(Z_i) + \varepsilon_{i+1}$, $i \geq 0$, where $(F^i)$ is a dynamical system and $(\varepsilon_i)$ is some additive and centered i.i.d. dynamical noise. In particular, consistency of two histogram-based methods is established if (a) $F:M \to M$ is continuous and $(\varepsilon_i)$ is bounded, or (b) $F$ is bounded and $\varepsilon_i$ is absolutely continuous. Note that the first case shows that in the absence of dynamical and observational noise there is a method which can learn to identify $F$ whenever it is continuous but otherwise unknown. However, it is unclear how to extend the methods of [32] to deal with observational noise.

Variants of the forecasting problem for general stationary ergodic processes $(Z_i)$ have been extensively studied in the literature. One often considered variant is *static autoregression* (see [22], page 569, and the references therein) where the goal is to find sequences $\hat{f}_m(Z_{-1}, \ldots, Z_{-m})$ of estimators that converge almost surely to $\mathbb{E}(Z_0 | Z_{-1}, \ldots, Z_{-\infty})$, which is known to be the least squares optimal one-step-ahead forecaster using an infinite past of observations. However, even if forecasters using a longer history of observations are considered in (2)–(4), the goal of static autoregression cannot be compared to our concept of consistency. Indeed, in static autoregression the goal is to find a near-optimal prediction for $\tilde{x}_0$ using the previously observed $\tilde{x}_{-1}, \ldots, \tilde{x}_{-m}$ of the *same* trajectory, whereas our goal is to use the observations to build a predictor which predicts near optimal for *arbitrary* future observations. In machine learning terminology, static autoregression is thus an "on-line" learning problem whereas our notion of consistency defines a "batch" learning problem.

Learning methods for estimating $\mathbb{E}(Z_0 | Z_{-1}, \ldots, Z_{-\infty})$ in a sense similar to (4) are considered by, for example, [29, 30]; unfortunately these methods require $\alpha$- or $\beta$-mixing conditions for $(Z_i)$ that cannot be satisfied by nontrivial dynamical systems. Finally, a result by Nobel [31] shows that there is no method that is universally consistent for classification and regression problems where the data is generated by an arbitrary stationary ergodic process $(Z_i)$. In particular this result shows that our general consistency Theorem 2.4 cannot be extended to such $(Z_i)$.

If the observational noise process $\mathcal{E}$ is mixing in the ergodic sense, then it is not hard to check that the process described by (1) is ergodic and hence it satisfies a strong law of large numbers by Birkhoff's theorem. Using the recent results in [39], we then see that there exists a support vector machine (see the next section for a description) *depending* on $F$ and $\mathcal{E}$ which is consistent in the sense of (4). However, [39] does not provide an explicit method for finding a consistent SVM even if both $F$ and $\mathcal{E}$ are known. Consequently, it is fair to say that though SVMs do not have principal limitations for the forecasting problem described by (1)–(4), there is currently no theoretically sound way to use them. The goal of this work is to address this issue by showing that certain SVMs are consistent for all pairs $(F, \mathcal{E})$ of Lipschitz continuous $F$ and bounded $\mathcal{E}$ that have a sufficiently fast decay of correlations for Lipschitz continuous functions. In particular, we show that these SVMs are consistent for all uniformly smooth expanding or hyperbolic dynamics $F$ and all bounded i.i.d. noise processes $\mathcal{E}$.

The rest of this work is organized as follows: In Section 1 we recall the definition of support vector machines (SVMs). Then, in Section 2, we present a consistency result for SVMs and general stochastic processes that have a sufficiently fast decay of correlations. This result is then applied to the above forecasting problem in Section 3, where we also briefly review some dynamical systems with a sufficiently fast decay of correlations. Possible future extensions of this work are discussed in Section 4. Finally, the proofs of the two main results can be found in Sections 5 and 6, respectively.

**1. Support vector machines.** The goal of this section is to briefly describe support vector machines, which were first introduced by [7, 15] as a method for learning binary classification tasks. Since then, they have been generalized to other problem domains such as regression and anomaly detection, and nowadays they are considered to be one of the state-of-the-art machine learning methods for these problem domains. For a thorough introduction to SVMs, we refer the reader to the books [16, 36, 42].

Let us begin by introducing some notation related to SVMs. To this end, let us fix two nonempty closed sets $X \subset \mathbb{R}^d$ and $Y \subset \mathbb{R}$, and a measurable function $L: X \times Y \times \mathbb{R} \to [0, \infty)$, which in the following is called loss function (note that this is a more general concept of a loss function than the informal notion of a loss function used in the introduction). For a finite sequence $T = ((x_1, y_1), \ldots, (x_n, y_n)) \in (X \times Y)^n$ and a function $f: X \to \mathbb{R}$, we define the empirical $L$-risk by

$$\mathcal{R}_{L,T}(f) := \frac{1}{n} \sum_{i=0}^{n-1} L(x_i, y_i, f(x_i)).$$

Moreover, for a distribution $P$ on $X \times Y$, we write

$$\mathcal{R}_{L,P}(f) := \int_{X \times Y} L(x, y, f(x)) \, dP(x, y)$$

and $\mathcal{R}^*_{L,P} := \inf\{\mathcal{R}_{L,p}(f) | f: \mathbb{R}^d \to \mathbb{R}^d \text{ measurable}\}$ for the $L$-risk and minimal $L$-risk associated to $P$. Now, let $H$ be the reproducing kernel Hilbert space (RKHS) of a measurable kernel $k: X \times X \to \mathcal{R}$ (see [1] for a general theory of such spaces). Given a finite sequence $T \in (X \times Y)^n$ and a regularization parameter $\lambda > 0$, support vector machines construct a function $f_{T,\lambda,H}: X \to \mathbb{R}$ satisfying

(5) $\quad \lambda \|f_{T,\lambda,H}\|^2_H + \mathcal{R}_{L,T}(f_{T,\lambda,H}) = \inf_{f \in H} (\lambda \|f\|^2_H + \mathcal{R}_{L,T}(f)).$

In the following we are mainly interested in the commonly used Gaussian RBF kernels $k_\sigma : X \times X \to \mathbb{R}$ defined by

$$k_\sigma(x, x') := \exp(-\sigma^2 \|x - x'\|_2^2), \qquad x, x' \in X,$$

where $X \subset \mathbb{R}^d$ is a nonempty subset and $\sigma > 0$ is a free parameter called the width. We write $H_\sigma(X)$ for the corresponding RKHSs, which are described in some detail in [40]. Finally, for SVMs using a Gaussian kernel $k_\sigma$, we write $f_{T,\lambda,\sigma} := f_{T,\lambda,H_\sigma(X)}$ in order to simplify notation.

It is well known that if $L$ is a *convex loss function* in the sense that $L(x, y, \cdot) : \mathbb{R} \to [0, \infty)$ is convex for all $(x, y) \in X \times Y$, then there exists a unique $f_{T,\lambda,H}$. Moreover, in this case (5) becomes a strictly convex optimization problem which can be solved by, for example, simple gradient descent algorithms. However, for specific losses, including the least squares loss, other more efficient algorithmic approaches are used in practice; see [36, 41, 42, 43]. Let us now introduce some additional properties of loss functions:

DEFINITION 1.1. A loss function $L : X \times Y \times \mathbb{R} \to [0, \infty)$ is called:

(i) *Differentiable* if $L(x, y, \cdot) : \mathbb{R} \to [0, \infty)$ is differentiable for all $(x, y) \in X \times Y$. In this case the derivative is denoted by $L'(x, y, \cdot)$.

(ii) *Locally Lipschitz continuous* if for all $a \geq 0$ there exists a constant $c_a \geq 0$ such that for all $x \in X$, $y \in Y$ and all $t, t' \in [-a, a]$ we have

$$|L(x, y, t) - L(x, y, t')| \leq c_a |t - t'|.$$

In this case the smallest possible constant $c_a$ is denoted by $|L|_{a,1}$.

(iii) *Lipschitz continuous* if $|L|_1 := \sup_{a \geq 0} |L|_{a,1} < \infty$.

With the help of these definitions we can now summarize assumptions on the loss function $L$ that we will use frequently.

ASSUMPTION L. The loss $L : X \times Y \times \mathbb{R} \to [0, \infty)$ is convex, differentiable and locally Lipschitz continuous in the above sense, and it also satisfies $L(x, y, 0) \leq 1$ for all $(x, y) \in X \times Y$. Moreover, for the derivative $L'$ there exists a constant $c \in [0, \infty)$ such that for all $(x, y, t), (x', y', t') \in X \times Y \times \mathbb{R}$ we have $|L'(x, y, 0)| \leq c$ and

(6) $\qquad |L'(x, y, t) - L'(x', y', t')| \leq c \|(x, y, t) - (x', y', t')\|_2.$

Note that combining the two assumptions on $L'$ yields $|L'(x, y, t)| \leq c(1 + |t|)$ for all $(x, y, t) \in X \times Y \times \mathbb{R}$, and from this it is not hard to conclude that $|L|_{a,1} \leq c(1 + a)$ for all $a > 0$.

Since the Assumption L is rather complex let us now illustrate it for two particular classes of loss functions used in many SVM variants.

EXAMPLE 1.2. A loss $L:X \times Y \times \mathbb{R} \to [0,\infty)$ of the form $L(x,y,t) = \varphi(yt)$ for a suitable function $\varphi:\mathbb{R} \to \mathbb{R}$ and all $x \in X$, $y \in Y := \{-1,1\}$ and $t \in \mathbb{R}$, is called *margin-based*. Obviously, $L$ is convex, continuous, (locally) Lipschitz continuous or differentiable if and only if $\varphi$ is. In addition, convexity of $L$ implies local Lipschitz continuity of $L$. Furthermore, recall that [6] showed that $L$ is suitable for binary classification tasks if and only if $\varphi$ is differentiable at 0 with $\varphi'(0) < 0$.

Let us now consider Assumption L. Obviously, the first part is satisfied if and only if $\varphi$ is convex and differentiable, and also satisfies $\varphi(0) \leq 1$. Note that the latter can always be ensured by rescaling $\varphi$. Furthermore, we have $L'(x,y,t) = y\varphi'(yt)$ and by considering the cases $y = y'$ and $y \neq y'$ separately we see that (6) is satisfied if and only if $\varphi'$ is Lipschitz continuous and satisfies

$$|\varphi'(t) + \varphi'(t')| \leq c(1 + |t + t'|), \qquad t, t' \in \mathbb{R},$$

for a constant $c > 0$. Finally, the condition $|L'(x,y,0)| = |\varphi'(0)| \leq c$ is always satisfied for sufficiently large $c$. From these considerations we conclude that the classical SVM losses $\varphi(t) = (1-t)_+$ and $\varphi(t) = (1-t)_+^2$, where $(x)_+ := \max\{0,x\}$, do *not* satisfy Assumption L, whereas the least square loss and the logistic loss defined by $\varphi(t) = (1-t)^2$ and $\varphi(t) = \ln(1+\exp(-t))$, respectively, fulfill Assumption L.

EXAMPLE 1.3. A loss $L:X \times Y \times \mathbb{R} \to [0,\infty)$ of the form $L(y,t) = \psi(y-t)$ for a suitable function $\psi:\mathbb{R} \to \mathbb{R}$ and all $x \in X$, $y \in Y \subset \mathbb{R}$ and $t \in \mathbb{R}$, is called *distance-based*. Recall that distance-based losses such as the least squares loss $\psi(r) = r^2$, Huber's insensitive loss $\psi(r) = \min\{r^2, \max\{1, 2|r| - 1\}\}$, the logistic loss $\psi(r) = \ln((1+e^r)^2 e^{-r}) - \ln 4$ or the $\varepsilon$-insensitive loss $\psi(r) = (|r| - \varepsilon)_+$ are usually used for regression.

In order to consider Assumption L we assume that $Y$ is a compact subset of $\mathbb{R}$. Then it is easy to see that the first part of Assumption L is satisfied if and only if $\psi$ is convex and differentiable, and also satisfies $\sup_{y \in Y} \psi(y) \leq 1$. Note that the latter can always be ensured by rescaling $\psi$ since the convexity of $\psi$ implies its continuity. Furthermore, we have $L'(x,y,t) = -\psi'(y-t)$, and hence we see that (6) is satisfied if and only if $\psi'$ is Lipschitz continuous. Finally, every convex and differentiable function is continuously differentiable and hence we can always ensure $|L'(x,y,0)| = |\psi'(y)| \leq c$. From these considerations we immediately see that all of the above distance-based losses besides the $\varepsilon$-insensitive loss satisfy Assumption L.

**2. Consistency of SVMs for a class of stochastic processes.** The goal of this section is to establish consistency of SVMs for a class of stochastic processes having a uniform decay of correlations for Lipschitz continuous functions. This result will then be used to establish consistency of SVMs for the forecasting problem and suitable combinations of dynamical systems $F$ and noise processes $\mathcal{E}$.

Let us begin with some notation. To this end, let us assume that we have a probability space $(\Omega, \mathcal{A}, \mu)$, a measurable space $(Z, \mathcal{B})$ and a measurable map $T : \Omega \to Z$. Then $\sigma(T)$ denotes the smallest $\sigma$-algebra on $\Omega$ for which $T$ is measurable. Moreover, $\mu_T$ denotes the $T$-image measure of $\mu$, which is defined by $\mu_T(B) := \mu(T^{-1}(B))$, $B \subset Z$ measurable. Recall that a *stochastic process* $\mathcal{Z} := (Z_n)_{n \geq 0}$, that is, a sequence of measurable maps $Z_n : \Omega \to Z$, $n \geq 0$, is called *identically distributed* if $\mu_{Z_n} = \mu_{Z_m}$ for all $n, m \geq 0$. In this case we write $P := \mu_{Z_0}$ in the following. Moreover, $\mathcal{Z}$ is called *second-order stationary* if $\mu_{(Z_{i_1+i}, Z_{i_2+i})} = \mu_{(Z_{i_1}, Z_{i_2})}$ for all $i_1, i_2, i \geq 1$, and it is said to be *stationary* if $\mu_{(Z_{i_1+i}, \ldots, Z_{i_n+i})} = \mu_{(Z_{i_1}, \ldots, Z_{i_n})}$ for all $n, i, i_1, \ldots, i_n \geq 1$.

The following definition introduces the *correlation sequence* for stochastic processes that will be used throughout this work.

DEFINITION 2.1. Let $(\Omega, \mathcal{A}, \mu)$ be a probability space, $(Z, \mathcal{B})$ be a measurable space, $\mathcal{Z}$ be a $Z$-valued, identically distributed process on $\Omega$ and $P := \mu_{Z_0}$. Then for $\psi, \varphi \in L_2(P)$ the $n$th correlation, $n \geq 0$, is defined by

$$\mathrm{cor}_{\mathcal{Z},n}(\psi, \varphi) := \int_\Omega \psi(Z_0) \cdot \varphi(Z_n) \, d\mu - \int_Z \psi \, dP \cdot \int_Z \varphi \, dP.$$

Obviously, if $\mathcal{Z}$ is an i.i.d. process, we have $\mathrm{cor}_{\mathcal{Z},n}(\psi, \varphi) = 0$ for all $\varphi, \psi \in L_2(P)$ and $n \geq 0$, and this remains true if $\psi \circ Z_0$ and $\varphi \circ Z_n$ are only uncorrelated. Consequently, if $\lim_{n \to \infty} \mathrm{cor}_{\mathcal{Z},n}(\psi, \varphi) = 0$ the corresponding speed of convergence provides information about how fast $\psi \circ Z_0$ becomes uncorrelated from $\varphi \circ Z_n$. This idea has been extensively used in the statistical literature in terms of, for example, the $\alpha$-mixing coefficients

$$\alpha(\mathcal{Z}, n) := \sup_{\substack{A \in \mathcal{F}_{-\infty}^0 \\ B \in \mathcal{F}_n^\infty}} |\mu(A \cap B) - \mu(A)\mu(B)|,$$

where $\mathcal{F}_i^j$ is the initial $\sigma$-algebra of $Z_i, \ldots, Z_j$. These and related (stronger) coefficients together with examples including, for example, certain Markov chains, ARMA processes, and GARCH processes are discussed in detail in the survey article [10] and the books [8, 11, 21]. Moreover, for processes $\mathcal{Z}$ satisfying $\alpha(\mathcal{Z}, n) \leq cn^{-\alpha}$ for some constant $c > 0$ and all $n \geq 1$ it was recently described in [39] how to find a regularization sequence $(\lambda_n)$ for which the corresponding SVM is consistent. Unfortunately, however, it is well known that every nontrivial ergodic dynamical system is *not* $\alpha$-mixing, that is, it does not satisfy $\lim_{n \to \infty} \alpha(\mathcal{Z}, n) = 0$, and therefore the result of [39] cannot be used to investigate consistency for the forecasting problem. On the other hand, various dynamical systems enjoy a uniform decay rate over smaller sets of functions such as Lipschitz continuous functions (see Section 3 for some examples). This leads to the following definition:

DEFINITION 2.2. Let $(\Omega, \mathcal{A}, \mu)$ be a probability space, $Z \subset \mathbb{R}^d$ be a compact set, $\mathcal{Z}$ be a $Z$-valued, identically distributed process on $\Omega$ and $P := \mu_{Z_0}$. Moreover, let $(\gamma_i)_{i \geq 0}$ be a strictly positive sequence converging to 0. Then $\mathcal{Z}$ is said to have a *decay of correlations of the order* $(\gamma_i)$ if for all $\psi, \varphi \in \mathrm{Lip}(Z)$ there exists a constant $\kappa_{\psi,\varphi} \in [0, \infty)$ such that

$$\tag{7} |\mathrm{cor}_{\mathcal{Z},i}(\psi, \varphi)| \leq \kappa_{\psi,\varphi} \gamma_i, \qquad i \geq 0,$$

where $\mathrm{Lip}(Z)$ denotes the set of all Lipschitz continuous $f : Z \to \mathbb{R}$.

Recall (see, e.g., Theorem 4.13 in Vol. 3 of [11]) that for every $Z$-valued, identically distributed process $\mathcal{Z}$ and all bounded functions $\psi, \varphi : Z \to \mathbb{R}$ we have

$$|\mathrm{cor}_{\mathcal{Z},i}(\psi, \varphi)| \leq 2\pi \|\psi\|_\infty \|\varphi\|_\infty \alpha(\mathcal{Z}, i), \qquad i \geq 1.$$

Since Lipschitz continuous functions on compacta are bounded, we hence see that $\alpha$-mixing processes have a decay of correlations of the order $(\alpha(\mathcal{Z}, i))$. In Section 3 we will present some examples of dynamical systems that are not $\alpha$-mixing but have a nontrivial decay of correlations.

Let us now summarize our assumptions on the process $\mathcal{Z}$ which we will make in the rest of this section.

ASSUMPTION Z. The process $\mathcal{Z} = (X_i, Y_i)_{i \geq 0}$ is defined on the probability space $(\Omega, \mathcal{A}, \mu)$ and is $X \times Y$-valued, where $X \subset \mathbb{R}^d$ and $Y \subset \mathbb{R}$ are compact subsets. Moreover $\mathcal{Z}$ is second-order stationary.

Finally, we will need the following mutually exclusive assumptions on the regularization sequence and the kernel width of SVMs:

ASSUMPTION S1. For a fixed strictly positive sequence $(\gamma_i)_{i \geq 0}$ converging to 0 and a locally Lipschitz continuous loss $L$ the monotone sequences $(\lambda_n) \subset (0, 1]$ and $(\sigma_n) \subset [1, \infty)$ satisfy $\lim_{n \to \infty} \lambda_n = 0$, $\sup_{n \geq 1} e^{-\sigma_n} |L|_{\lambda_n^{-1/2}, 1} < \infty$,

$$\sup_{n \geq 1} \frac{\lambda_n \sigma_n^{4d}}{|L|_{\lambda_n^{-1/2}, 1}} < \infty \quad \text{and} \quad \lim_{n \to \infty} \frac{|L|^3_{\lambda_n^{-1/2}, 1} \sigma_n^2}{n \lambda_n^4} \sum_{i=0}^{n-1} \gamma_i = 0.$$

ASSUMPTION S2. For a fixed strictly positive sequence $(\gamma_i)_{i \geq 0}$ converging to 0 and a locally Lipschitz continuous loss $L$ the sequences $(\lambda_n) \subset (0, 1]$ and $(\sigma_n) \subset [1, \infty)$ satisfy $\lim_{n \to \infty} \lambda_n \sigma_n^d = 0$,

$$\lim_{n \to \infty} \frac{\lambda_n \sigma_n^{4d}}{|L|_{\lambda_n^{-1/2}, 1}} = \infty \quad \text{and} \quad \lim_{n \to \infty} \frac{\sigma_n^{2+12d}}{n \lambda_n} \sum_{i=0}^{n-1} \gamma_i = 0.$$

ASSUMPTION S3. For a fixed strictly positive sequence $(\gamma_i)_{i\geq 0}$ converging to 0 and a locally Lipschitz continuous loss $L$ the monotone sequences $(\lambda_n) \subset (0,1]$ and $(\sigma_n) \subset [1,\infty)$ satisfy $\lim_{n\to\infty} \lambda_n = 0$, $\lim_{n\to\infty} e^{-\sigma_n}|L|_{\lambda_n^{-1/2},1} = \infty$,

$$\sup_{n\geq 1} \frac{\lambda_n \sigma_n^{4d}}{|L|_{\lambda_n^{-1/2},1}} < \infty \quad \text{and} \quad \lim_{n\to\infty} \frac{|L|^6_{\lambda_n^{-1/2},1} e^{-2\sigma_n}}{n\lambda_n^4} \sum_{i=0}^{n-1} \gamma_i = 0.$$

REMARK 2.3. In order to illustrate the Assumptions S1, S2 and S3, let us assume for simplicity that $L$ is Lipschitz continuous; the case that $L$ is the least squares loss will be considered in Remark 3.4. Now note that for Lipschitz continuous losses Assumption S3 cannot be satisfied and hence it suffices to consider Assumptions S1 and S2.

Let us first assume $\sum_{i\geq 0} \gamma_i < \infty$ as well as $\lambda_n := n^{-\alpha}$ and $\sigma_n := n^\beta$ for $n \geq 1$ and constants $\alpha > 0$ and $\beta \geq 0$. Then Assumption S1 is met if $\alpha \geq 4d\beta$ and $4\alpha + 2\beta < 1$, whereas Assumption S2 is met if $d\beta < \alpha < 4d\beta$ and $\alpha + (2+12d)\beta < 1$. In particular, for $\beta = 0$ Assumption S1 is met if $0 < \alpha < 1/4$, whereas Assumption S2 cannot be met in this case.

Finally, we consider a milder assumption on the decay of correlations, namely $n^{-1}\sum_{i=0}^{n-1} \gamma_i \leq c(1+\ln n)^{-1}$, for a constant $c > 0$ and all $n \geq 1$. Obviously, this is satisfied if we assume that $(\gamma_i)$ has some arbitrary polynomial decay. Let us consider the sequences $\lambda_n := (1+\ln n)^{-\alpha}$ and $\sigma_n := (1+\ln n)^\beta$ for $n \geq 1$ and $\alpha > 0$ and $\beta \geq 0$. Then Assumption S1 is met if $\alpha \geq 4d\beta$ and $4\alpha + 2\beta < 1$, whereas Assumption S2 is met if $d\beta < \alpha < 4d\beta$ and $\alpha + (2+12d)\beta < 1$. In particular, for $\beta = 0$ Assumption S1 is met if $0 < \alpha < 1/4$, while Assumption S2 cannot be met.

The illustrations above show that both Assumptions S1 and S2 consist of two contrary conditions, namely one which implies that $\lambda_n$ tends to 0 and another one which ensures that this speed is not too fast. Roughly speaking, the first condition guarantees that the approximation error tends to zero (see Lemma 5.4), but since this simultaneously means that the statistical error becomes larger, the second condition is needed to ensure that the latter error still tends to zero (see the proof of Theorem 2.4). This trade-off between approximation and statistical error is typical for consistent learning algorithms (see the books [19] and [22] for several such examples).

With the help of these assumptions we can now establish the announced consistency of SVMs.

THEOREM 2.4. *Let $\mathcal{Z} = (X_i, Y_i)_{i\geq 0}$ be a stochastic process satisfying Assumption Z. We write $P := \mu_{(X_0,Y_0)}$ and assume that $\mathcal{Z}$ has a decay of correlations of some order $(\gamma_i)$. In addition, let $L: X \times Y \times \mathbb{R} \to [0,\infty)$ be* a loss satisfying Assumption L. Then for all sequences $(\lambda_n) \subset (0,1]$ and $(\sigma_n) \subset [1,\infty)$ satisfying Assumptions S1, S2 or S3 and all $\varepsilon \in (0,1]$ we have

$$\lim_{n \to \infty} \mu(\omega \in \Omega : |\mathcal{R}_{L,P}(f_{T_n(\omega),\lambda_n,\sigma_n}) - \mathcal{R}_{L,P}^*| > \varepsilon) = 0,$$

where $T_n(\omega) := ((X_0(\omega), Y_0(\omega)), \ldots, (X_{n-1}(\omega), Y_{n-1}(\omega)))$ and $f_{T_n(\omega),\lambda_n,\sigma_n}$ is the SVM forecaster defined by (5).

Theorem 2.4 in particular applies to stochastic processes that are $\alpha$-mixing with rate $(\gamma_i)$. However, the Assumptions S1, S2 and S3 ensuring consistency are substantially stronger than the ones obtained in [39] for such processes. On the other hand, there are interesting stochastic processes that are not $\alpha$-mixing but still enjoy a reasonably fast decay of correlations. Since we are mainly interested in the forecasting problem we will delay the discussion of such examples to the next section.

**3. Consistency of SVMs for the forecasting problem.** In this section we present our main result, which establishes the consistency of SVMs for the forecasting problem described by (1)–(4) if the dynamical system enjoys a certain decay of correlations. In addition, we discuss some examples of such systems.

We begin by first revisiting our informal problem description given in the introduction. To this end, let $M \subset \mathbb{R}^d$ be a compact set and $F : M \to M$ be a map such that the dynamical system $\mathcal{D} := (F^i)_{i \geq 0}$ has a unique ergodic measure $\mu$. Moreover, let $\mathcal{E} = (\varepsilon_i)_{i \geq 0}$ be a $\mathbb{R}^d$-valued stochastic process which is (stochastically) independent of $\mathcal{D}$. Then the process that generates the noisy observations (1) is $(F^i + \varepsilon_i)_{i \geq 0}$. In particular, a sequence of observations $(\tilde{x}_0, \ldots, \tilde{x}_n)$ generated by this process is of the form (1) for a conjoint initial state. Now recall that, given an observation of the system at some arbitrary time, our goal is to forecast the next *observable* state. Consequently, we will use the training set

$$\begin{aligned}(8) \quad T_n(x,\varepsilon) &:= ((\tilde{x}_0, \tilde{x}_1), \ldots, (\tilde{x}_{n-1}, \tilde{x}_n)) \\ &= ((x + \varepsilon_0, F(x) + \varepsilon_1), \ldots, (F^{n-1}(x) + \varepsilon_{n-1}, F^n(x) + \varepsilon_n))\end{aligned}$$

whose input/output pairs are consecutive observable states. Now note that a single sample $(F^{i-1}(x) + \varepsilon_{i-1}, F^i(x) + \varepsilon_i)$ depends on the pair $(\varepsilon_i, \varepsilon_{i+1})$ and thus we have to consider the process of such pairs. The following assumption summarizes the needed requirements of the process $\mathcal{N} := ((\varepsilon_i, \varepsilon_{i+1}))_{i \geq 0}$.

ASSUMPTION N. For the $\mathbb{R}^{2d}$-valued stochastic process $\mathcal{N}$ there exist a constant $B > 0$ and a probability measure $\nu$ on $[-B, B]^{d\mathbb{N}_0}$ such that the coordinate process $\mathcal{E} := (\pi_0 \circ S^i)_{i \geq 0}$ is stationary with respect to $\nu$ and satisfies $\mathcal{N} = (\pi_0 \circ S^i, \pi_0 \circ S^{i+1})_{i \geq 0}$, where $S$ denotes the shift operator $(x_i)_{i \geq 0} \mapsto (x_{i+1})_{i \geq 0}$ and $\pi_0$ denotes the projection $(x_i)_{i \geq 0} \mapsto x_0$.

Before we state our main result we note that the input variable $x + \varepsilon$ and the output variable $F(x) + \varepsilon'$ are $d$-dimensional vectors. Consequently, our notion of a loss introduced in Section 1 needs a refinement which captures the ideas of the introduction. To this end we state the following assumption:

ASSUMPTION LD. For the function $L: \mathbb{R}^d \to [0, \infty)$ there exists a distance-based loss satisfying Assumption L such that its representing function $\psi: \mathbb{R}^d \to [0, \infty)$ has a unique global minimum at 0 and satisfies

(9) $\quad L(r_1, \ldots, r_d) = \psi(r_1) + \cdots + \psi(r_d), \qquad (r_1, \ldots, r_d) \in \mathbb{R}^d.$

Obviously, if $L$ satisfies Assumption LD, then $L$ is a loss in the sense of the introduction. Moreover note that the specific form (9) makes it possible to consider the coordinates of the output variable *separately.* Consequently, we will use the forecaster

(10) $\quad \bar{f}_{T, \lambda, \sigma} := (f_{T^{(1)}, \lambda, \sigma}, \ldots, f_{T^{(d)}, \lambda, \sigma}),$

where $f_{T^{(j)}, \lambda, \sigma}$ is the SVM solution obtained by considering the distance-based loss defined by $\psi$ and $T^{(j)} := ((\tilde{x}_0, \pi_j(\tilde{x}_1)), \ldots, (\tilde{x}_{n-1}, \pi_j(\tilde{x}_n)))$ which is obtained by projecting the output variable of $T$ onto its $j$th-coordinate via the coordinate projection $\pi_j : \mathbb{R}^d \to \mathbb{R}$. In other words, we build the forecaster $\bar{f}_{T, \lambda, \sigma}$ by training $d$ different SVMs on the training sets $T^{(1)}, \ldots, T^{(d)}$.

With the help of these preparations we can now present our main result, which establishes consistency for such a forecaster.

THEOREM 3.1. *Let $M \subset \mathbb{R}^d$ be a compact set, $F: M \to M$ be a Lipschitz continuous map such that the dynamical system $\mathcal{D} := (F^i)_{i \geq 0}$ has a unique ergodic measure $\mu$, and $\mathcal{N}$ be a stochastic process satisfying Assumption N. Assume that both processes $\mathcal{D}$ and $\mathcal{N}$ have a decay of correlations of the order $(\gamma_i)$. Moreover, let $L: \mathbb{R}^d \to [0, \infty)$ be a function satisfying Assumption LD. Then for all sequences $(\lambda_n) \subset (0, 1]$ and $(\sigma_n) \subset [1, \infty)$ satisfying Assumptions S1, S2 or S3 and all $\varepsilon \in (0, 1]$ we have*

$$\lim_{n \to \infty} \mu \otimes \nu((x, \varepsilon) \in M \times [-B, B]^{d\mathbb{N}} : |\mathcal{R}_{L, P}(\bar{f}_{T_n(x, \varepsilon), \lambda_n, \sigma_n}) - \mathcal{R}^*_{L, P}| > \varepsilon) = 0,$$

*where $T_n(x, \varepsilon)$ is defined by (8) and the risks are given by (2) and (3).*

Note that if $\mathcal{E}$ is an i.i.d. process, then $\mathcal{N}$ has a decay of correlations of any order. Moreover, if $\mathcal{E}$ is $\alpha$-mixing with mixing rate $(\gamma_i)$, then $\mathcal{N}$ has a decay of correlations of order $(\gamma_i)$. Finally, if $\mathcal{D}$ has a decay of correlations $(\gamma_i')$ and $\mathcal{N}$ has a decay of correlations $(\gamma_i'')$, then they obviously both have a decay of correlations $(\gamma_i)$, where $\gamma_i := \max\{\gamma_i', \gamma_i''\}$. In particular, noise processes having slowly decaying correlations will slow down learning even though the system $\mathcal{D}$ may have a fast decay of correlations.

Let us now discuss some examples of classes of dynamical systems enjoying at least a polynomial decay of correlations. Since the existing literature on such systems is vast these examples are only meant to be illustrations for situations where Theorem 3.1 can be applied and are *not* intended to provide an overview of known results. However, compilations of known results can be found in the survey articles [3, 28] and the book [2].

EXAMPLE 3.2 (*Smooth expanding dynamics*). Let $M$ be a compact connected Riemannian manifold and $F: M \to M$ be $C^{1+\varepsilon}$ for some $\varepsilon > 0$. Furthermore assume that there exist constants $c > 0$ and $\lambda > 1$ such that

$$\max\{\|DF_x^n(v)\| : x \in M, v \in T_xM \text{ with } \|v\| = 1\} \geq c\lambda^n$$

for all $n \geq 0$, where $T_xM$ denotes the tangent space of $M$ at $x$ and $DF_x^n$ denotes the derivative of $F^n$ at $x$. Then it is a classical result that $F$ possesses a unique ergodic measure which is absolutely continuous with respect to the Riemannian volume. Moreover, it is well known (see, e.g., [33] and the references mentioned in [28], Theorem 5) that there exists a $\tau > 0$ such that the dynamical system has decay of correlations of the order $(e^{-\tau i})$. Generalizations of this result to piecewise smooth and piecewise (non)-uniformly expanding dynamics are discussed in [3]. Finally, [28], Theorem 10, recalls results (together with references) for non-uniformly expanding dynamics having either exponential or polynomial decay of correlations.

EXAMPLE 3.3 (*Smooth hyperbolic dynamics*). If $F$ is a topologically mixing $C^{1+\varepsilon}$ Anosov or Axiom A diffeomorphism, then it is well known (see, e.g., [9, 34]) that there exists a $\tau > 0$ such that the dynamical system has a decay of correlations of the order $(e^{-\tau i})$. Moreover, Baldi [3] lists various extensions of this result to, for example, smooth nonuniformly hyperbolic systems and hyperbolic systems with singularities.

Besides these classical results and their extensions, Baldi [3] also compiles a list of "parabolic" or "intermittent" systems having a polynomial decay.

Let us now consider the forecasting problem for the least squares loss. To this end we first observe that the function $L(r) := \|r\|_2^2$, $r \in \mathbb{R}^d$, satisfies Assumption LD since the least squares loss satisfies Assumption L as we have discussed in Example 1.3. Let us now additionally assume that the noise is *pairwise independent* (i.e., $\varepsilon_i$ and $\varepsilon_{i'}$ are independent if $i \neq i'$) and centered [i.e., it satisfies $\mathbb{E}_{\varepsilon \sim \nu} \pi_0(\varepsilon) = 0$]. For a forecaster $f = (f_1, \ldots, f_d): \mathbb{R}^d \to \mathbb{R}^d$ we then obtain

$$\mathcal{R}_{L,P}(f) = \iint \sum_{j=1}^d (\pi_j(F(x) + \varepsilon_1) - f_j(x + \varepsilon_0))^2 \nu(d\varepsilon) \mu(dx)$$

$$= \iint \sum_{j=1}^{d} (\pi_j(F(x)) - f_j(x+\varepsilon_0))^2 \nu(d\varepsilon)\mu(dx) + \int \|\varepsilon_0\|_2^2 \nu(d\varepsilon)$$

$$=: \overline{\mathcal{R}}_{L,P}(f) + \int \|\varepsilon_0\|_2^2 \nu(d\varepsilon),$$

where $\pi_j : \mathbb{R}^d \to \mathbb{R}$ denotes the $j$th coordinate projection. Consequently, a forecaster $f$ that approximately minimizes the $L$-risk is also an approximate forecaster of the *true* next state in the sense of $\overline{\mathcal{R}}_{L,P}(\cdot)$. Before we combine this observation with Theorem 3.1 let us first rephrase Assumptions S1, S2 and S3 for the least squares loss.

ASSUMPTION S1-LS. For a strictly positive sequence $(\gamma_i)_{i\geq 0}$ converging to 0 the monotone sequences $(\lambda_n) \subset (0,1]$ and $(\sigma_n) \subset [1,\infty)$ satisfy $\lim_{n\to\infty} \lambda_n = 0$, $\sup_{n\geq 1} e^{-\sigma_n} \lambda_n^{-1/2} < \infty$,

$$\sup_{n\geq 1} \lambda_n \sigma_n^{8d/3} < \infty \quad \text{and} \quad \lim_{n\to\infty} \frac{\sigma_n^2}{n\lambda_n^{11/2}} \sum_{i=0}^{n-1} \gamma_i = 0.$$

ASSUMPTION S2-LS. For a strictly positive sequence $(\gamma_i)_{i\geq 0}$ converging to 0 the sequences $(\lambda_n) \subset (0,1]$ and $(\sigma_n) \subset [1,\infty)$ satisfy $\lim_{n\to\infty} \lambda_n \sigma_n^d = 0$,

$$\lim_{n\to\infty} \lambda_n \sigma_n^{8d/3} = \infty \quad \text{and} \quad \lim_{n\to\infty} \frac{\sigma_n^{2+12d}}{n\lambda_n} \sum_{i=0}^{n-1} \gamma_i = 0.$$

ASSUMPTION S3-LS. For a strictly positive sequence $(\gamma_i)_{i\geq 0}$ converging to 0 the monotone sequences $(\lambda_n) \subset (0,1]$ and $(\sigma_n) \subset [1,\infty)$ satisfy $\lim_{n\to\infty} \lambda_n = 0$, $\lim_{n\to\infty} e^{-\sigma_n} \lambda_n^{-1/2} = \infty$,

$$\sup_{n\geq 1} \lambda_n \sigma_n^{8d/3} < \infty \quad \text{and} \quad \lim_{n\to\infty} \frac{e^{-\sigma_n}}{n\lambda_n^7} \sum_{i=0}^{n-1} \gamma_i = 0.$$

REMARK 3.4. In order to illustrate the Assumptions S1-LS, S2-LS and S3-LS, let us first assume $\sum_{i\geq 0} \gamma_i < \infty$ as well as $\lambda_n := n^{-\alpha}$ and $\sigma_n := n^\beta$ for $n \geq 1$ and constants $\alpha > 0$ and $\beta \geq 0$. Then Assumption S1-LS is met if $3\alpha \geq 8d\beta > 0$ and $11\alpha + 4\beta < 2$, whereas Assumption S2-LS is met if $\alpha + (2+12d)\beta < 1$ and $d\beta < \alpha < \frac{8}{3}d\beta$. Finally Assumption S3-LS is satisfied if $\beta = 0$ and $0 < \alpha < 1/7$.

Let us now consider the milder assumption $n^{-1}\sum_{i=0}^{n-1} \gamma_i \leq c(1+\ln n)^{-1}$ which has already been considered in Remark 2.3 for Lipschitz continuous losses. To this end, we again consider the sequences $\lambda_n := (1+\ln n)^{-\alpha}$ and $\sigma_n := (1+\ln n)^\beta$ for $n \geq 1$ and constants $\alpha > 0$ and $\beta \geq 0$. Then Assumption S1-LS is met if $3\alpha \geq 8d\beta > 0$ and $11\alpha + 4\beta < 2$, whereas Assumption S2-LS is met if $\alpha + (2+12d)\beta < 1$ and $d\beta < \alpha < \frac{8}{3}d\beta$. Finally Assumption S3-LS is satisfied if $\beta = 0$ and $0 < \alpha < 1/7$.

With these preparations we can now state a result showing that SVMs using a least squares loss can be used to forecast the next *true* state of the dynamical system if the observational noise is sufficiently benign.

COROLLARY 3.5. *Let $M \subset \mathbb{R}^d$ be a compact set and $F:M \to M$ be a Lipschitz continuous map such that the dynamical system $\mathcal{D} := (F^i)_{i \geq 0}$ has a unique ergodic measure $\mu$. Moreover, let $\mathcal{E} = (\varepsilon_i)_{i \geq 0}$ be an i.i.d. process of $[-B,B]^d$-valued and centered random variables. Assume that $\mathcal{D}$ has a decay of correlations of the order $(\gamma_i)$. Moreover, let $L:\mathbb{R}^d \to [0,\infty)$ be defined by $L(r) := \|r\|_2^2$, $r \in \mathbb{R}^d$. Then for all sequences $(\lambda_n) \subset (0,1]$ and $(\sigma_n) \subset [1,\infty)$ satisfying Assumptions S1-LS, S2-LS or S3-LS and all $\varepsilon \in (0,1]$ we have*

$$\lim_{n \to \infty} \mu \otimes \nu((x,\varepsilon) \in M \times [-B,B]^{d\mathbb{N}} : |\overline{\mathcal{R}}_{L,P}(\bar{f}_{T_n(x,\varepsilon),\lambda_n,\sigma_n}) - \overline{\mathcal{R}}_{L,P}^*| > \varepsilon) = 0,$$

*where $\overline{\mathcal{R}}_{L,P}^* := \inf\{\overline{\mathcal{R}}_{L,P}(f) | f : \mathbb{R}^d \to \mathbb{R}^d \text{ measurable}\}$.*

It is interesting to note that the above corollary does *not* require the noise to be symmetric. Instead it only requires centered noise, that is, the observations are not systematically biased in a certain direction.

Let us end this section with the following remark that rephrases Theorem 3.1 and its corollary for situations with summable decays of correlations.

REMARK 3.6 (*Universal consistency*). If the sequence $(\gamma_i)$ bounding the correlation is *summable*, that is, $\sum \gamma_i < \infty$, then the Assumptions S1, S2, S3, S1-LS, S2-LS and S3-LS are *independent* of both the dynamical system and the observational noise process. Consequently, using sequences satisfying one of these assumptions yields an SVM which is consistent for *all* such pairs of dynamical systems and observational noise processes. In other words, such an SVM can learn the optimal forecaster without knowing specifics of the dynamical systems and the observational noise. To be a bit more specific, let us assume, for example, that we use the least squares loss and sequences $\lambda_n := n^{-\alpha}$ and $\sigma_n := n^\beta$, $n \geq 1$, for fixed $\alpha$ and $\beta$ satisfying $3\alpha \geq 8d\beta > 0$ and $11\alpha + 4\beta < 2$. Then the corresponding SVM is consistent for all bounded observational noise processes having a summable $\alpha$-mixing rate and all ergodic dynamical systems on $M$ which are defined by a Lipschitz continuous $F:M \to M$ and have a summable decay of correlations. Note that this class of dynamical systems includes, but is not limited to, smooth uniformly expanding or hyperbolic dynamics. Finally, if the noise process is also i.i.d. and centered then this SVM actually learns to forecast the next *true* state.

It is interesting to note that a similar consistency result holds for all noise processes having a polynomial decay of $\alpha$-mixing coefficients and all ergodic dynamical systems on $M$ which are defined by a Lipschitz continuous $F: M \to M$ and have a polynomial decay of correlations. Indeed, for such combinations SVMs using sequences $\lambda_n := (1+\ln n)^{-\alpha}$ and $\sigma_n := (1+\ln n)^{\beta}$ with, for example, fixed $\alpha$ and $\beta$ satisfying $3\alpha \geq 8d\beta > 0$ and $11\alpha + 4\beta < 2$ are consistent.

**4. Discussion.** The goal of this work was to show that, in principle, support vector machines can learn how to predict one-step-ahead noisy observation of a dynamical system without knowing specifics of the dynamical system or the observational noise besides a certain, rather general stochasticity. However, there remain several open questions which can be subject to further research:

*More general losses and kernels.* In the statistical part of our analysis, we used an approach which is based on a "stability" argument. However, it is also possible to use a "skeleton" argument based on covering numbers, instead. Utilizing the latter, it seems possible to relax the assumptions on the loss $L$ by making stronger assumptions on both $(\lambda_n)$ and $(\sigma_n)$. A particular loss which is interesting in this direction would be the $\varepsilon$-insensitive loss used in classical SVMs for regression. Another possible extension of our work is considering different kernels, such as the kernels that generate Sobolev spaces. In fact, we only focused on Gaussian RBF kernels since these kernels are the most commonly used in practice.

*Learning rates.* So far we have only shown that the risk of the SVM solution converges to the smallest possible risk. However, for practical considerations the *speed* of this convergence is of great importance, too. The proof we utilized already gives such learning rates if a *quantitative* version of the Approximation Lemma 5.4 is available, which is possible if, for example, quantitative assumptions on the smoothness of $F$ and the regularity of $\nu$ are made. However, since we conjecture that the statistical part of our analysis is not sharp we have not presented a corresponding result. In this regard we note that recently [14] established a concentration result for piecewise regular expanding and topologically mixing maps of the interval $[0,1]$, which is substantially stronger than our elementary Chebyshev inequality of Lemma 5.8. We believe that such a concentration result can be used to substantially sharpen the statistical part of our analysis.

*Perturbed dynamics.* Another extension of the current work is to consider systems that are perturbed by some noise. Our general consistency result in Theorem 2.4 suggests that such an extension is possible whenever the perturbed system has a decay of correlations. In this regard we note that for some perturbed systems of expanding maps the decay of correlations has already been bounded in [5], and it would be interesting to investigate whether they can be used to prove consistency of SVMs.

*Longer past.* So far, we have only used the present observation to forecast the next observation, but it is not hard to see that in almost any system/noise combination the minimal risk $\mathcal{R}_{L,P}^*$ reduces if one uses additional past observations. On the other hand it appears that the learning problem becomes harder in this case since we have to approximate a function which lives on a higher dimensional input space, and hence there seems to be a trade-off for finite sample sizes. While investigating this trade-off in more detail seems to be possible with the techniques developed in this work, we again assume that the statistical part of our analysis is not sharp enough to obtain a meaningful picture of this trade-off.

**5. Proof of Theorem 2.4.** The goal of this section is to prove Theorem 2.4. Since the proof requires several preliminary results, we divided this section into subsections, which provide these prerequisites.

5.1. *Some basics on the decay of correlations.* The main goal of this section is to establish some *uniform* bounds on the sequence of correlations.

Let us begin introducing some notation. To this end, we fix a probability space $(\Omega, \mathcal{A}, \mu)$, a measurable space $(Z, \mathcal{B})$ and a $Z$-valued, identically distributed process on $\Omega$. For $P := \mu_{Z_0}$ and $\psi, \varphi \in L_2(P)$ we then write $\operatorname{cor}_{\mathcal{Z}}(\psi, \varphi) := (\operatorname{cor}_{\mathcal{Z},n}(\psi, \varphi))_{n \geq 0}$ for the sequence of correlations of $\psi$ and $\varphi$. Clearly, this gives a bilinear map $\operatorname{cor}_{\mathcal{Z}} : L_2(P) \times L_2(P) \to \ell_\infty$, which in the following is called the *correlation operator.* The following key theorem, which goes back to an unpublished note [13] of Collet (see also page 101 in [4]), can be used to establish continuity of the correlation operator. Before we present this result let us first recall that a Banach space $E$ is said to be continuously embedded into the Banach space $F$ if $E \subset F$ and the natural inclusion map $\operatorname{id} : E \to F$ is continuous.

THEOREM 5.1. *Let $(\Omega, \mathcal{A}, \mu)$ be a probability space, $(Z, \mathcal{B})$ be a measurable space, $\mathcal{Z}$ be a $Z$-valued, identically distributed process on $\Omega$ and $P := \mu_{Z_0}$. Moreover, let $E_1$ and $E_2$ be Banach spaces that are continuously embedded into $L_2(P)$ and let $F$ be a Banach space that is continuously embedded into $\ell_\infty$. If for all $\psi \in E_1$ and all $\varphi \in E_2$ the correlation operator satisfies*

$$\operatorname{cor}_{\mathcal{Z}}(\psi, \varphi) \in F,$$

*then there exists a constant $c \in [0, \infty)$ such that*

$$\|\operatorname{cor}_{\mathcal{Z}}(\psi, \varphi)\|_F \leq c \cdot \|\psi\|_{E_1} \|\varphi\|_{E_2}, \qquad \psi \in E_1, \varphi \in E_2.$$

For the sake of completeness the proof of this key result can be found in the Appendix. The most obvious examples of Banach spaces $F$ in the above theorem are the spaces $\ell_p$. However, in the literature on dynamical systems results on the sequence of correlations are usually stated in the form

$$|\mathrm{cor}_{\mathcal{Z},n}(\psi,\varphi)| \leq \kappa_{\psi,\varphi}\gamma_n, \qquad n \geq 0,$$

where $(\gamma_n)$ is a strictly positive sequence converging to 0 and $\kappa_{\psi,\varphi}$ is a constant depending on $\psi$ and $\varphi$. To apply Theorem 5.1 in this situation we obviously need Banach spaces which capture such a behavior of $\mathrm{cor}_{\mathcal{Z}}(\cdot,\cdot)$. Therefore, let us fix a strictly positive sequence $\gamma := (\gamma_n)_{n \geq 0}$ such that $\lim_{n \to \infty} \gamma_n = 0$. For a sequence $b := (b_n) \subset \mathbb{R}$ we define

$$\|b\|_{\Lambda(\gamma)} := \sup_{n \geq 0} \frac{|b_n|}{\gamma_n}.$$

Moreover, we write

$$\Lambda(\gamma) := \{(b_n) \subset \mathbb{R} : \|(b_n)\|_{\Lambda(\gamma)} < \infty\}.$$

The following lemma establishes some basic properties of $(\Lambda(\gamma), \|\cdot\|_{\Lambda(\gamma)})$.

LEMMA 5.2. *The pair $(\Lambda(\gamma), \|\cdot\|_{\Lambda(\gamma)})$ is a Banach space continuously embedded into $\ell_\infty$ and we have $\|\mathrm{id} : \Lambda(\gamma) \to \ell_\infty\| \leq \|\gamma\|_\infty$.*

PROOF. The fact that $(\Lambda(\gamma), \|\cdot\|_{\Lambda(\gamma)})$ is a normed space is elementary to prove. Moreover, we have

$$\|b\|_{\Lambda(\gamma)} = \sup_{n \geq 0} \frac{|b_n|}{\gamma_n} \geq \sup_{n \geq 0} \frac{|b_n|}{\|\gamma\|_\infty} = \frac{\|b\|_\infty}{\|\gamma\|_\infty},$$

and hence we find $\|\mathrm{id} : \Lambda(\gamma) \to \ell_\infty\| \leq \|\gamma\|_\infty$. Finally, let $(b^{(i)})_{i \geq 1}$ be a Cauchy sequence in $\Lambda(\gamma)$. The previous step shows that it is also a Cauchy sequence in $\ell_\infty$, and by the completeness of $\ell_\infty$ there consequently exists a sequence $b := (b_n) \in \ell_\infty$ such that $\lim_{i \to \infty} \|b^{(i)} - b\|_\infty = 0$. Let us now fix an $\varepsilon > 0$. Then there exists an index $i_0 \geq 0$ such that for all $i, j \geq i_0$ we have $\|b^{(i)} - b^{(j)}\|_{\Lambda(\gamma)} \leq \varepsilon$. Consequently, for fixed $N \geq 0$ we have

$$\sup_{n=0,\ldots,N} \frac{|b_n^{(i)} - b_n^{(j)}|}{\gamma_n} \leq \|b^{(i)} - b^{(j)}\|_{\Lambda(\gamma)} \leq \varepsilon,$$

and by taking the limit $j \to \infty$ we conclude

$$\sup_{n=0,\ldots,N} \frac{|b_n^{(i)} - b_n|}{\gamma_n} \leq \varepsilon.$$

However, $N$ was arbitrary and hence we find $\|b^{(i)} - b\|_{\Lambda(\gamma)} \leq \varepsilon$ for all $i \geq i_0$. In other words we have shown that $(b^{(i)})_{i \geq 1}$ converges to $b$ in $\|\cdot\|_{\Lambda(\gamma)}$. □

Combining the above lemma with Theorem 5.1 we immediately obtain the following corollary:

COROLLARY 5.3. *Let $(\Omega, \mathcal{A}, \mu)$ be a probability space, $(Z, \mathcal{B})$ be a measurable space, $\mathcal{Z}$ be a $Z$-valued, identically distributed process on $\Omega$ and $P := \mu_{Z_0}$. Moreover, let $E_1$ and $E_2$ be Banach spaces that are continuously embedded into $L_2(P)$. In addition, let $\gamma := (\gamma_n)_{n \geq 0}$ be a strictly positive sequence such that $\lim_{n \to \infty} \gamma_n = 0$. If for all $\psi \in E_1$ and all $\varphi \in E_2$ there exists a constant $\kappa_{\psi,\varphi} \in [0, \infty)$ such that*

$$|\operatorname{cor}_{\mathcal{Z},n}(\psi, \varphi)| \leq \kappa_{\psi,\varphi} \gamma_n$$

*for all $n \geq 0$, then there exists a constant $c \in [0, \infty)$ such that*

$$|\operatorname{cor}_{\mathcal{Z},n}(\psi, \varphi)| \leq c \|\psi\|_{E_1} \cdot \|\varphi\|_{E_2} \cdot \gamma_n, \qquad \psi \in E_1, \varphi \in E_2, n \geq 0.$$

5.2. *Some properties of Gaussian RBF kernels.* In this subsection we establish some properties of Gaussian RBF kernels which will be heavily used in the proof of Theorem 2.4. Let us begin with an approximation result.

LEMMA 5.4. *Let $X \subset \mathbb{R}^d$ and $Y \subset \mathbb{R}$ be compact subsets, $L : X \times Y \times \mathbb{R} \to [0, \infty)$ be a convex locally Lipschitz continuous loss and $P$ be a probability measure on $X \times Y$ such that $\mathcal{R}_{L,P}(0) < \infty$. Then for all sequences $(\lambda_n) \subset (0, 1]$ and $(\sigma_n) \subset [1, \infty)$ satisfying*

$$\lim_{n \to \infty} \lambda_n \sigma_n^d = 0, \tag{11}$$

*we have*

$$\lim_{n \to \infty} \left( \inf_{f \in H_{\sigma_n}(X)} \lambda_n \|f\|_{H_{\sigma_n}(X)}^2 + \mathcal{R}_{L,P}(f) \right) = \mathcal{R}_{L,P}^*.$$

PROOF. For $\sigma > 0$ we write $\mathcal{R}_{L,P,H_\sigma(X)}^* := \inf\{\mathcal{R}_{L,P}(f) : f \in H_\sigma(X)\}$. Since $L$ is locally Lipschitz continuous and $\mathcal{R}_{L,P}(0) < \infty$, the discussion after (4) in [38] shows that it is a $P$-integrable Nemitski loss in the sense of [38]. Now recall (see [37]) that $H_\sigma(X)$ is universal, that is, it is dense in $C(X)$, and hence [38], Corollary 1, shows $\mathcal{R}_{L,P,H_\sigma(X)}^* = \mathcal{R}_{L,P}^*$ for all $\sigma > 0$. Let us now fix an $\varepsilon > 0$. The above discussion then shows that there exists an $f_\varepsilon \in H_1(X)$ such that $\mathcal{R}_{L,P}(f_\varepsilon) \leq \mathcal{R}_{L,P}^* + \varepsilon$. Furthermore, by (11) there exists an $n_0 \geq 0$ such that

$$\lambda_n \sigma_n^d \leq \varepsilon \|f_\varepsilon\|_{H_1(X)}^{-2}, \qquad n \geq n_0.$$

Since $\sigma_n \geq 1$ we also know $f_\varepsilon \in H_{\sigma_n}(X)$ and $\|f_\varepsilon\|_{H_{\sigma_n}(X)}^2 \leq \sigma_n^d \|f_\varepsilon\|_{H_1(X)}^2$ by [40], Corollary 6, and therefore we obtain

$$\inf_{f \in H_{\sigma_n}(X)} \lambda_n \|f\|_{H_{\sigma_n}(X)}^2 + \mathcal{R}_{L,P}(f) \leq \lambda_n \|f_\varepsilon\|_{H_{\sigma_n}(X)}^2 + \mathcal{R}_{L,P}(f_\varepsilon) \leq \mathcal{R}_{L,P}^* + 2\varepsilon$$

for all $n \geq n_0$. From this we easily deduce the assertion. $\square$

Before we establish the next result let us recall that a function $f: X \to \mathbb{R}$ on a subset $X \subset \mathbb{R}^d$ is called Lipschitz continuous if there exists a constant $c \in [0, \infty)$ such that $|f(x) - f(x')| \leq c\|x - x'\|_2$ for all $x, x' \in X$. In the following the smallest such constant is denoted by $|f|_1$ and the set of all Lipschitz continuous functions is denoted by $\mathrm{Lip}(X)$. Moreover, recall that if $X$ is compact then $\mathrm{Lip}(X)$ together with the norm $\|f\|_{\mathrm{Lip}(X)} := \max\{\|f\|_\infty, |f|_1\}$ forms a Banach space. In this case $\mathrm{Lip}(X)$ is also closed under multiplication. Indeed, for $f, g \in \mathrm{Lip}(X)$ and $x, x' \in X$ we have

$$|f(x)g(x) - f(x')g(x')| \leq \|f\|_\infty \cdot |g|_1 |x - x'| + |f|_1 \cdot \|g\|_\infty |x - x'|$$

and hence we obtain $fg \in \mathrm{Lip}(X)$ with $|fg|_1 \leq \|f\|_\infty \cdot |g|_1 + |f|_1 \cdot \|g\|_\infty$. Our next result shows that every function in $H_\sigma(X)$ is Lipschitz continuous.

LEMMA 5.5. *Let $X \subset \mathbb{R}^d$ be a nonempty set and $\sigma > 0$. Then every $f \in H_\sigma(X)$ is Lipschitz continuous with $|f|_1 \leq \sqrt{2}\sigma \|f\|_{H_\sigma(X)}$.*

PROOF. Let us write $\Phi: X \to H_\sigma(X)$ for the canonical feature map defined by $\Phi(x) := k_\sigma(x, \cdot)$. Now recall that $\Phi$ satisfies the reproducing property

$$f(x) = \langle \Phi(x), f \rangle, \qquad x \in X, f \in H_\sigma(X),$$

and hence in particular $k_\sigma(x', x) = \langle \Phi(x), \Phi(x') \rangle$ for all $x, x' \in X$. Using these equalities together with $1 - e^{-t} \leq t$ for $t \geq 0$ we obtain

$$\begin{aligned}|f(x) - f(x')| &= |\langle \Phi(x) - \Phi(x'), f \rangle| \\ &\leq \|f\|_{H_\sigma(X)} \cdot \|\Phi(x) - \Phi(x')\|_{H_\sigma(X)} \\ &= \|f\|_{H_\sigma(X)} \sqrt{\langle \Phi(x), \Phi(x) \rangle + \langle \Phi(x'), \Phi(x') \rangle - 2\langle \Phi(x), \Phi(x') \rangle} \\ &= \|f\|_{H_\sigma(X)} \sqrt{2 - 2\exp(-\sigma^2 \|x - x'\|_2^2)} \\ &\leq \sqrt{2}\sigma \|f\|_{H_\sigma(X)} \|x - x'\|_2,\end{aligned}$$

that is, we have proved the assertion. $\square$

In the following we consider certain orthonormal bases (ONBs) of $H_\sigma(X)$. To this end, let us first recall that in [40], Theorem 5, it was shown that $(e_n)_{n \geq 0}$, where $e_n: \mathbb{R} \to \mathbb{R}$ is defined by

$$(12) \qquad e_n(x) := \sqrt{\frac{2^n \sigma^{2n}}{n!}} x^n e^{-\sigma^2 x^2}, \qquad x \in \mathbb{R},$$

forms an ONB of $H_\sigma(\mathbb{R})$. Moreover, it was shown that if $X \subset \mathbb{R}$ has a nonempty interior, the restrictions of $e_n$ to $X$ form an ONB of $H_\sigma(X)$.

The following lemma establishes upper bounds on $\|e_n\|_\infty$ if $X$ is a closed interval.

LEMMA 5.6.  *Let $\sigma > 0$ and $a > 0$ be fixed real numbers and $(e_n)_{n\geq 0}$ be the ONB of $H_\sigma([-a,a])$, where $e_n$ is defined by the restriction of (12) to $[-a,a]$. Then we have $\|e_n\|_\infty \leq (2\pi n)^{-1/4}$ for all $n \geq 1$ and*

$$\|e_n\|_\infty \leq \sqrt{\frac{2^n a^{2n} \sigma^{2n}}{n!}} e^{-a^2\sigma^2} \tag{13}$$

*for all $n \geq 2a^2\sigma^2$. In addition, for $n \geq 8ea^2\sigma^2$ we have*

$$\left(\sum_{i=n+1}^\infty \|e_i\|_\infty^2\right)^{1/2} \leq \left(\frac{2}{\pi(n+1)}\right)^{1/4} 2^{-(n+1)} e^{-a^2\sigma^2}, \tag{14}$$

*and for $a\sigma \geq 1$ we also have*

$$\left(\sum_{i=0}^\infty \|e_i\|_\infty^2\right)^{1/2} \leq \sqrt{6a\sigma}. \tag{15}$$

PROOF. Elementary calculus shows

$$e_n'(x) = \sqrt{\frac{2^n \sigma^{2n}}{n!}} x^{n-1} e^{-\sigma^2 x^2} (n - 2\sigma^2 x^2)$$

for all $n \geq 1$ and $x \in \mathbb{R}$. From this we conclude $e_n'(x^*) = 0$ if and only if $x^* = \pm\sqrt{\frac{n}{2\sigma^2}}$ or $x^* = 0$. Therefore it is not hard to see that the function defined in (12) attains its global extrema at $x^* = \pm\sqrt{\frac{n}{2\sigma^2}}$, and hence we obtain

$$\|e_n\|_\infty \leq \sqrt{\frac{n^n}{n!}} e^{-n/2} \leq \sqrt{\frac{n^n}{\sqrt{2\pi n}\, n^n e^{-n}}} e^{-n/2} = (2\pi n)^{-1/4}$$

for all $n \geq 1$ by Stirling's formula. Moreover, $n \geq 2a^2\sigma^2$ implies $|x^*| \geq a$ and, in this case, it is not hard to see that the function $|e_n|$ actually attains its maximum at $\pm a$. From these considerations we conclude (13).

For the proof of (14) we recall that the remainder of the Taylor series of the exponential function satisfies

$$\sum_{i=n+1}^\infty \frac{y^i}{i!} \leq 2\frac{|y|^{n+1}}{(n+1)!}$$

for $|y| \leq 1 + n/2$. Since $n \geq 8ea^2\sigma^2$ implies $2a^2\sigma^2 \leq 1 + n/2$, we consequently obtain

$$\sum_{i=n+1}^{\infty} \|e_i\|_\infty^2 \leq \sum_{i=n+1}^{\infty} \frac{2^i a^{2i} \sigma^{2i}}{i!} e^{-2a^2\sigma^2} \leq \frac{2^{n+2} a^{2(n+1)} \sigma^{2(n+1)}}{(n+1)!} e^{-2a^2\sigma^2}$$

$$\leq 2 \frac{2^{n+1} a^{2(n+1)} \sigma^{2(n+1)} e^{(n+1)}}{\sqrt{2\pi(n+1)}(n+1)^{(n+1)}} e^{-2a^2\sigma^2}$$

$$\leq \left(\frac{2}{\pi(n+1)}\right)^{1/2} 4^{-(n+1)} e^{-2a^2\sigma^2}.$$

From this we easily deduce (14). Finally, for the proof of (15), we observe

$$\sum_{i=0}^{\lceil 8ea^2\sigma^2 \rceil} \|e_i\|_\infty^2 \leq 1 + (2\pi)^{-1/2} + \sum_{i=2}^{\lceil 8ea^2\sigma^2 \rceil} (2\pi i)^{-1/2}$$

$$\leq 1 + (2\pi)^{-1/2} + (2\pi)^{-1/2} \int_1^{8ea^2\sigma^2 + 1} x^{-1/2}\, dx$$

$$\leq 1 + (2\pi)^{-1/2} + (e/\pi)^{-1/2} 4a\sigma$$

$$\leq 3/2 + 4a\sigma.$$

Combining this estimate with (14), we then obtain

$$\sum_{i=0}^{\infty} \|e_i\|_\infty^2 = \sum_{i=0}^{\lceil 8ea^2\sigma^2 \rceil} \|e_i\|_\infty^2 + \sum_{i=\lceil 8ea^2\sigma^2 \rceil + 1}^{\infty} \|e_i\|_\infty^2$$

$$\leq 3/2 + 4a\sigma + \left(\frac{2}{\pi(\lceil 8ea^2\sigma^2 \rceil + 1)}\right)^{1/2} 4^{-(\lceil 8ea^2\sigma^2 \rceil + 1)} e^{-2a^2\sigma^2}$$

$$\leq 3/2 + 4a\sigma + \left(\frac{1}{8e\pi a^2 \sigma^2}\right)^{1/2} 4^{-8ea^2\sigma^2} e^{-2a^2\sigma^2}$$

$$\leq 2 + 4a\sigma,$$

and from the latter we easily obtain (15). □

Our next goal is to generalize the above result to the multi-dimensional case. To this end, recall that the tensor product $f \otimes g : X \times X \to \mathbb{R}$ of two functions $f, g : X \to \mathbb{R}$ is defined by $f \otimes g(x, x') := f(x)g(x')$, $x, x' \in X$. Obviously, for bounded functions we have $\|f \otimes g\|_\infty = \|f\|_\infty \|g\|_\infty$.

For a multi-index $\eta = (n_1, \ldots, n_d) \in \mathbb{N}_0^d$ we use the notation $\eta \geq n$ if $n_i \geq n$ for all $i = 1, \ldots, d$. Moreover, we write

(16) $$e_\eta := e_{n_1} \otimes \cdots \otimes e_{n_d}, \qquad \eta = (n_1, \ldots, n_d) \in \mathbb{N}_0^d,$$

where $e_{n_i}$ is defined by (12). Then [40], Theorem 5, shows that $(e_\eta)_{\eta \in \mathbb{N}_0^d}$ is an ONB of $H_\sigma(\mathbb{R}^d)$ and the restrictions of the members of this ONB to $[-a,a]^d$ form an ONB of $H_\sigma([-a,a]^d)$. The following lemma generalizes the estimates of Lemma 5.6 to this multi-dimensional ONB.

COROLLARY 5.7. *For $\sigma > 0$ and $a > 0$ satisfying $a\sigma \geq 1$ and $d \in \mathbb{N}$, let $(e_\eta)_{\eta \in \mathbb{N}_0^d}$ be the restriction of the ONB (16) to $[-a,a]^d$. Then for $n \geq 8ea^2\sigma^2$ we have*

$$\left( \sum_{\substack{\eta \in \mathbb{N}_0^d \\ \exists i: \eta_i > n}} \|e_\eta\|_\infty^2 \right)^{1/2} \leq \sqrt{d} e^{-a^2\sigma^2} (6a\sigma)^{(d-1)/2} \left( \frac{2}{\pi(n+1)} \right)^{1/4} 2^{-(n+1)}.$$

PROOF. Using $\|e_{i_1} \otimes \cdots \otimes e_{i_d}\|_\infty = \|e_{i_1}\|_\infty \cdots \|e_{i_d}\|_\infty$ we obtain

$$\sum_{\substack{\eta \in \mathbb{N}_0^d \\ \exists i: \eta_i > n}} \|e_\eta\|_\infty^2 \leq d \sum_{i_1=n+1}^\infty \sum_{i_2=0}^\infty \cdots \sum_{i_d=0}^\infty \prod_{j=1}^d \|e_{i_j}\|_\infty^2$$

$$= d \left( \sum_{i=n+1}^\infty \|e_i\|_\infty^2 \right) \left( \sum_{i=0}^\infty \|e_i\|_\infty^2 \right)^{d-1}$$

$$\leq d \left( \frac{2}{\pi(n+1)} \right)^{1/2} 2^{-2(n+1)} e^{-2a^2\sigma^2} (6a\sigma)^{d-1}$$

by Lemma 5.6. From this we immediately obtain the assertion. □

5.3. *A concentration inequality in RKHSs.* In this subsection we will establish a concentration inequality for RKHS-valued functions and for processes which have a certain decay of correlations. This concentration result will then be the key ingredient in the statistical analysis of the proof of Theorem 2.4.

Let us begin by recalling a simple inequality that will be used several times:

LEMMA 5.8. *Let $\mathcal{Z} = (Z_i)_{i \geq 0}$ be a second-order stationary $Z$-valued process on $(\Omega, \mathcal{A}, \mu)$. Then for $P := \mu_{Z_0}$, $f \in L_2(P)$, $n \geq 1$ and $\delta > 0$ we have*

$$(17) \quad \mu\left( \omega \in \Omega : \left| \frac{1}{n} \sum_{i=0}^{n-1} f \circ Z_i(\omega) - \mathbb{E}_P f \right| \geq \delta \right) \leq \frac{2}{n\delta^2} \sum_{i=0}^{n-1} \mathrm{cor}_{\mathcal{Z},i}(f,f).$$

For the following results we have to introduce more notation: Given a bounded measurable kernel $k: X \times X \to \mathbb{R}$ with RKHS $H$, we write $\Phi: X \to$ $H$, $\Phi(x) := k(x, \cdot)$ for the canonical feature map. Moreover, for a bounded measurable function $h: X \times Y \to \mathbb{R}$ and a distribution $P$ on $X \times Y$ we write $\mathbb{E}_P h\Phi$ for the Bochner integral (see [20]) of the $H$-valued function $h\Phi$. Similarly, given $T := ((x_1, y_1), \ldots, (x_n, y_n)) \in (X \times Y)^n$ we write

$$(18) \qquad \mathbb{E}_T h\Phi := \frac{1}{n} \sum_{i=1}^{n} h\Phi(x_i, y_i)$$

for the empirical counterpart of $\mathbb{E}_P h\Phi$. In order to motivate the following results we further mention that the proof of Theorem 2.4 will heavily rely on the estimate

$$\|f_{P,\lambda,H} - f_{T,\lambda,H}\|_H \leq \frac{1}{\lambda} \|\mathbb{E}_P h_\lambda \Phi - \mathbb{E}_T h_\lambda \Phi\|_H,$$

where $f_{P,\lambda,H}$ is the SVM solution (see Theorem 5.12 for an exact definition) one obtains by replacing the empirical risk $\mathcal{R}_{L,T}(\cdot)$ with the true risk $\mathcal{R}_{L,P}(\cdot)$ in (5) and $h_\lambda$ is a function independent of the training set $T$. Consequently, our next goal is to estimate terms of the form $\|\mathbb{E}_P h\Phi - \mathbb{E}_T h\Phi\|_H$. To this end, we begin with the following lemma which, roughly speaking, will be used to reduce RKHS-valued functions to $\mathbb{R}$-valued functions.

LEMMA 5.9. *Let $H$ be the separable RKHS of a bounded measurable kernel $k: X \times X \to \mathbb{R}$, let $\Phi: X \to H$ be the corresponding canonical feature map and $(e_i)_{i \geq 0}$ be an ONB of $H$. Moreover, let $Y$ be another measurable space, $P$ and $Q$ be probability measures on $X \times Y$ and $h \in L_1(P) \cap L_1(Q)$. Then for all $n \geq 0$ we have*

$$\|\mathbb{E}_P h\Phi - \mathbb{E}_Q h\Phi\|_H$$
$$\leq \left( \sum_{i=0}^{n} |\mathbb{E}_P he_i - \mathbb{E}_Q he_i|^2 \right)^{1/2} + \left( \sum_{i=n+1}^{\infty} \|e_i\|_\infty^2 \right)^{1/2} (\mathbb{E}_P |h| + \mathbb{E}_Q |h|).$$

PROOF. Let us define $S_n: H \to H$ by $\sum_{i \geq 0} \langle f, e_i \rangle e_i \mapsto \sum_{i=0}^{n} \langle f, e_i \rangle e_i$. Then we have

$$\|S_n \Phi(x) - \Phi(x)\|_H^2 = \left\| \sum_{i=n+1}^{\infty} \langle \Phi(x), e_i \rangle e_i \right\|_H^2$$
$$= \sum_{i=n+1}^{\infty} |\langle \Phi(x), e_i \rangle|^2$$
$$= \sum_{i=n+1}^{\infty} |e_i(x)|^2$$

by the reproducing property and hence we obtain

$$\|\mathbb{E}_P h\Phi - \mathbb{E}_Q h\Phi\|_H$$
$$\leq \|\mathbb{E}_P h\Phi - \mathbb{E}_P h S_n\Phi\|_H + \|\mathbb{E}_P h S_n\Phi - \mathbb{E}_Q h S_n\Phi\|_H$$
$$\quad + \|\mathbb{E}_Q h S_n\Phi - \mathbb{E}_Q h\Phi\|_H$$
$$\leq \mathbb{E}_P |h|\|\Phi - S_n\Phi\|_H + \|\mathbb{E}_P h S_n\Phi - \mathbb{E}_Q h S_n\Phi\|_H$$
$$\quad + \mathbb{E}_Q |h|\|\Phi - S_n\Phi\|_H$$
$$\leq \|\mathbb{E}_P h S_n\Phi - \mathbb{E}_Q h S_n\Phi\|_H + \left(\sum_{i=n+1}^{\infty} \|e_i\|_{\infty}^2\right)^{1/2}$$
$$\quad \times (\mathbb{E}_P |h| + \mathbb{E}_Q |h|).$$

Moreover, using the reproducing property we have $\langle \mathbb{E}_P h\Phi, e_i\rangle = \mathbb{E}_P h e_i$ and $\langle \mathbb{E}_Q h\Phi, e_i\rangle = \mathbb{E}_Q h e_i$, and thus we conclude

$$\|\mathbb{E}_P h S_n\Phi - \mathbb{E}_Q h S_n\Phi\|_H^2 = \sum_{i=0}^{n} |\langle \mathbb{E}_P h\Phi - \mathbb{E}_Q h\Phi, e_i\rangle|^2 = \sum_{i=0}^{n} |\mathbb{E}_P h e_i - \mathbb{E}_Q h e_i|^2.$$

Combining this equality with the previous estimate, we find the assertion. □

Before we can establish the concentration inequality for RKHS-valued functions, we finally need the following simple lemma.

LEMMA 5.10. *For $d \geq 1$ and $t > 18\, d \ln(d)$ we have $t^{-1/4} 2^{-t} \leq t^{-2d}$.*

PROOF. Obviously, it suffices to show

(19) $$t \ln 2 + (1/4 - 2d) \ln t \geq 0.$$

Let us first prove the case $d = 1$. Then (19) reduces to the assertion $h(t) := t \ln 2 - \frac{7}{4} \ln t \geq 0$. To establish the latter, note that we have $h'(t) = \ln 2 - \frac{7}{4} t^{-1}$ and hence $h'(t^*) = 0$ holds if and only if $t^* = \frac{7}{4 \ln 2}$. Simple considerations then show that $h$ has its only global minimum at $t^*$ and therefore we have $h(t) \geq h(t^*) \geq \frac{7}{4} - \frac{7}{4} \ln(\frac{7}{\ln 16}) > 0$.

Let us now consider the case $d \geq 2$. To this end we fix a $t > 18\, d \ln(d)$. Then there exists a unique $x > 18$ with $t = x\, d \ln(d)$, and hence we obtain

$$t \ln 2 + (1/4 - 2d) \ln t = x\, d \ln(d) \ln 2 + 1/4 \ln(x\, d \ln(d)) - 2d \ln(x\, d \ln(d))$$
$$> x\, d \ln(d) \ln 2 - 2d \ln(x\, d \ln(d))$$
$$= d(x \ln(d) \ln 2 - 2 \ln x - 2 \ln d - 2 \ln(\ln(d)))$$
$$> d\left(x \ln(d) \ln 2 - 2 \frac{\ln d}{\ln 2} \ln x - 2 \ln d - 2 \ln d\right)$$

$$= d\ln(d)\left(x\ln 2 - \frac{2}{\ln 2}\ln x - 4\right),$$

where in the last estimate we used $d \geq 2$. Now it is elementary to check that $x \mapsto x\ln 2 - \frac{2}{\ln 2}\ln x - 4$ is increasing on $[2(\ln 2)^{-2}, \infty)$ and since $18\ln 2 - \frac{2}{\ln 2}\ln 18 - 4 > 0$, we then obtain (19). □

THEOREM 5.11. *For $\sigma > 0$ and $a > 0$ satisfying $a\sigma \geq 1$ and $d \geq 1$, let $\Phi : [-a,a]^d \to H_\sigma([-a,a]^d)$ be the canonical feature map of the Gaussian RBF kernel and let $(e_\eta)_{\eta \in \mathbb{N}_0^d}$ be the ONB of $H_\sigma([-a,a]^d)$ which is considered in Corollary 5.7. In addition, let $Y$ be a measurable space and let $\mathcal{Z} = (X_i, Y_i)_{i \geq 0}$ be a $[-a,a]^d \times Y$-valued process on $(\Omega, \mathcal{A}, \mu)$ that is second-order stationary. Furthermore, let $(\gamma_i)_{i \geq 0}$ be a strictly positive sequence converging to zero, $h : [-a,a]^d \times Y \to \mathbb{R}$ be a bounded measurable function and $K_h \in [1, \infty)$ be a constant such that*

(20) $$\operatorname{cor}_{\mathcal{Z},i}(he_\eta, he_\eta) \leq K_h \gamma_i$$

*for all $i \geq 0$, $\eta \in \mathbb{N}_0^d$. Then for all $\varepsilon > 0$ satisfying both $\varepsilon \leq (1 + 8ea^2\sigma^2)^{-2d}$ and $\varepsilon \leq (18d\ln d)^{-2d}$ and all $n \geq 1$ we have*

$$\mu(\omega \in \Omega : \|\mathbb{E}_P h\Phi - \mathbb{E}_{T_n(\omega)} h\Phi\|_H \leq \varepsilon)$$
$$\geq 1 - \frac{2(1 + (1/(8ea^2\sigma^2))^d K_h C_{a\sigma,d,h}^3}{n\varepsilon^3} \sum_{i=0}^{n-1} \gamma_i,$$

*where $\mathbb{E}_{T_n(\omega)} h\Phi$ denotes the empirical Bochner integral (18) with respect to the data set $T_n(\omega) := (Z_0(\omega), \ldots, Z_{n-1}(\omega))$, and*

$$C_{a\sigma,d,h} := \left(1 + \frac{1}{8ea^2\sigma^2}\right)^{d/2}$$
$$+ 2\sqrt{d} e^{-a^2\sigma^2} (6a\sigma)^{(d-1)/2} \|h\|_\infty.$$

PROOF. Let us write

$$\delta := \left(\frac{\varepsilon}{C_{a\sigma,d,h}}\right)^{5/4}.$$

Using $C_{a\sigma,d,h} \geq (1 + \frac{1}{8ea^2\sigma^2})^{d/2} \geq 1$ and $\varepsilon \leq (1 + 8ea^2\sigma^2)^{-2d}$ we then find $\delta \leq (1 + 8ea^2\sigma^2)^{-5d/2}$, and consequently, there exists a natural number $m \geq 8ea^2\sigma^2$ such that $(m+1)^{-5d/2} \leq \delta < m^{-5d/2}$. Moreover, for later use we note that using $C_{a\sigma,d,h} \geq 1$ and $\varepsilon \leq (18d\ln d)^{-2d}$ yields $\delta^{-2/5d} \geq 18d\ln d$. Let us now consider an $\omega \in \Omega$ such that

(21) $$|\mathbb{E}_P he_\eta - \mathbb{E}_{T_n(\omega)} he_\eta| < \delta$$

for all $\eta \in \{0, \ldots, m\}^d$. By Lemma 5.9 and Corollary 5.7 we then obtain

$$\|\mathbb{E}_P h\Phi - \mathbb{E}_{T_n(\omega)} h\Phi\|_H$$

$$\leq \left(\sum_{\eta \leq m} |\mathbb{E}_P he_\eta - \mathbb{E}_{T_n(\omega)} he_\eta|^2\right)^{1/2}$$

$$+ \left(\sum_{\substack{\eta \in \mathbb{N}_0^d \\ \exists i: \eta_i > m}} \|e_\eta\|_\infty^2\right)^{1/2} (\mathbb{E}_P|h| + \mathbb{E}_{T_n(\omega)}|h|)$$

$$\leq (m+1)^{d/2} \delta + 2\sqrt{d} e^{-a^2\sigma^2} (6a\sigma)^{(d-1)/2} \left(\frac{2}{\pi(m+1)}\right)^{1/4} 2^{-(m+1)} \|h\|_\infty$$

$$\leq \left(1 + \frac{1}{8ea^2\sigma^2}\right)^{d/2} \delta^{4/5} + 2\sqrt{d} e^{-a^2\sigma^2} (6a\sigma)^{(d-1)/2} \delta^{1/(10d)} 2^{-\delta^{-2/(5d)}} \|h\|_\infty,$$

where in the last step we used the inequalities $8ea^2\sigma^2 \leq m < \delta^{-2/(5d)} \leq m+1$. Using Lemma 5.10 for $t := \delta^{-2/(5d)}$ we consequently obtain

$$\|\mathbb{E}_P h\Phi - \mathbb{E}_{T_n(\omega)} h\Phi\|_H$$

$$\leq \left(\left(1 + \frac{1}{8ea^2\sigma^2}\right)^{d/2} + 2\sqrt{d} e^{-a^2\sigma^2} (6a\sigma)^{(d-1)/2} \|h\|_\infty\right) \delta^{4/5} = \varepsilon.$$

Moreover, by Lemma 5.8 and a simple union bound argument we see that the probability of $\omega$ satisfying (21) for all $\eta \in \{0, \ldots, m\}^d$ simultaneously is not smaller than

$$1 - \sum_{\eta \in \{0,\ldots,m\}^d} \frac{2}{n\delta^2} \sum_{i=0}^{n-1} \operatorname{cor}_{\mathcal{Z},i}(he_\eta, he_\eta).$$

In addition, we have

$$\sum_{\eta \in \{0,\ldots,m\}^d} \frac{2}{n\delta^2} \sum_{i=0}^{n-1} \operatorname{cor}_{\mathcal{Z},i}(he_\eta, he_\eta) \leq \frac{2(m+1)^d}{n\delta^2} \sum_{i=0}^{n-1} K_h \gamma_i,$$

and since $8ea^2\sigma^2 \leq m < \delta^{-2/(5d)}$ we further estimate

$$\frac{2(m+1)^d}{n\delta^2} \leq \frac{2(1 + 1/(8ea^2\sigma^2))^d m^d}{n\delta^2} \leq \frac{2(1 + 1/(8ea^2\sigma^2))^d}{n\delta^{12/5}}$$

$$= \frac{2(1 + 1/(8ea^2\sigma^2))^d C_{a\sigma,d,h}^3}{n\varepsilon^3}.$$

Combining these estimates we then obtain the assertion. □

5.4. *Proof of Theorem 2.4.* For the proof of Theorem 2.4 we need some final preparations. Let us begin with the following result on the existence and uniqueness of infinite sample SVMs which is a slight extension of similar results established in [12, 18]:

THEOREM 5.12. *Let $L: X \times Y \times \mathbb{R} \to [0, \infty)$ be a convex, locally Lipschitz continuous loss function satisfying $L(x, y, 0) \leq 1$ for all $(x, y) \in (X \times Y)$, and let $P$ be a distribution on $X \times Y$. Furthermore, let $H$ be a RKHS of a bounded measurable kernel over $X$. Then for all $\lambda > 0$ there exists exactly one element $f_{P,\lambda,H} \in H$ such that*

$$(22) \qquad \lambda \|f_{P,\lambda,H}\|_H^2 + \mathcal{R}_{L,P}(f_{P,\lambda,H}) = \inf_{f \in H} \lambda \|f\|_H^2 + \mathcal{R}_{L,P}(f).$$

*Furthermore, we have $\|f_{P,\lambda,H}\|_H \leq \lambda^{-1/2}$.*

Note that the above theorem in particular yields $\|f_{T,\lambda,H}\|_H \leq \lambda^{-1/2}$ by considering the empirical measure associated to a training set $T \in (X \times Y)^n$. The following result which was (essentially) shown in [12, 18] describes the stability of the empirical SVM solutions.

THEOREM 5.13. *Let $X$ be a separable metric space, $L: X \times Y \times \mathbb{R} \to [0, \infty)$ be a convex, locally Lipschitz continuous loss function satisfying $L(x, y, 0) \leq 1$ for all $(x, y) \in (X \times Y)$ and let $P$ be a distribution on $X \times Y$. Furthermore, let $H$ be the RKHS of a bounded continuous kernel $k: X \times X \to \mathbb{R}$ and let $\Phi: X \to H$ be the corresponding canonical feature map. Then for all $\lambda > 0$ the function $h_\lambda: X \times Y \to \mathbb{R}$ defined by*

$$(23) \qquad h_\lambda(x, y) := L'(x, y, f_{P,\lambda}(x)), \qquad (x, y) \in X \times Y,$$

*is bounded and satisfies and*

$$(24) \quad \|f_{P,\lambda,H} - f_{T,\lambda,H}\|_H \leq \frac{1}{\lambda} \|\mathbb{E}_P h_\lambda \Phi - \mathbb{E}_T h_\lambda \Phi\|_H, \qquad T \in (X \times Y)^n.$$

PROOF OF THEOREM 2.4. Obviously, it suffices to consider sets $X$ of the form $X = [-a, a]^d$ for some $a \geq 1$. For $\sigma > 0$ and $\lambda > 0$ we write $h_{\lambda,\sigma}$ for the function we obtain by Theorem 5.13 for $H := H_\sigma(X)$. By the local Lipschitz continuity of $L$, $\|k_\sigma\|_\infty \leq 1$, Theorem 5.12 and (24) we then have

$$(25) \qquad \begin{aligned} &|\mathcal{R}_{L,P}(f_{T,\lambda,\sigma}) - \mathcal{R}_{L,P}(f_{P,\lambda,\sigma})| \\ &\qquad \leq \frac{|L|_{\lambda^{-1/2},1}}{\lambda} \|\mathbb{E}_P h_{\lambda,\sigma} \Phi - \mathbb{E}_T h_{\lambda,\sigma} \Phi\|_{H_\sigma(X)} \end{aligned}$$

for all $\sigma > 0$, $\lambda > 0$ and all $T \in (X \times Y)^n$. Moreover, using (23), (6) and Lemma 5.5 we have

$$\begin{aligned}
|h_{\lambda,\sigma}(x,y) &- h_{\lambda,\sigma}(x',y')| \\
&= |L'(x,y,f_{P,\lambda,\sigma}(x)) - L'(x',y',f_{P,\lambda,\sigma}(x'))| \\
&\leq c \cdot (|x-x'|^2 + |y-y'|^2 + |f_{P,\lambda,\sigma}(x) - f_{P,\lambda,\sigma}(x')|^2)^{1/2} \\
&\leq c \cdot (|x-x'|^2 + |y-y'|^2 + 2\sigma^2 \|f_{P,\lambda,\sigma}\|_{H_\sigma(X)}^2 |x-x'|^2)^{1/2} \\
&\leq 2c\sigma\lambda^{-1/2} \|(x,y) - (x',y')\|_2
\end{aligned}$$

for all $\sigma \geq 1$, $\lambda \in (0,1]$ and all $(x,y), (x',y') \in X \times Y$. Consequently, we find $|h_{\lambda,\sigma}|_1 \leq 2c\sigma\lambda^{-1/2}$. Moreover, we have

$$(26) \quad |h_{\lambda,\sigma}(x,y)| = |L'(x,y,f_{P,\lambda,\sigma}(x))| \leq \sup_{|t| \leq \lambda^{-1/2}} |L'(x,y,t)| \leq |L|_{\lambda^{-1/2},1}$$

for all $\lambda > 0$ and all $(x,y) \in X \times Y$. Let us now write $e_\eta^{(\sigma)}$ for the $\eta$th element, $\eta \in \mathbb{N}_0^d$, of the ONB of $H_\sigma(X)$ considered in Corollary 5.7. Combining the above estimates with Lemma 5.5 and the trivial bound $\|e_\eta^{(\sigma)}\|_\infty \leq \|e_\eta^{(\sigma)}\|_{H_\sigma(X)} \leq 1$ we obtain

$$|h_{\lambda,\sigma} e_\eta^{(\sigma)}|_1 \leq |h_{\lambda,\sigma}|_1 \|e_\eta^{(\sigma)}\|_\infty + \|h_{\lambda,\sigma}\|_\infty |e_\eta^{(\sigma)}|_1 \leq 5c\sigma\lambda^{-1/2}$$

for all $\lambda \in (0,1]$ and $\sigma \geq 1$, where in the last step we used the estimate $|L|_{a,1} \leq c(1+a)$, $a > 0$, which we derived after Assumption L. Since we further have $\|h_{\lambda,\sigma} e_\eta^{(\sigma)}\|_\infty \leq \|h_{\lambda,\sigma}\|_\infty \leq 2c\lambda^{-1/2}$, we find $\|h_{\lambda,\sigma} e_\eta^{(\sigma)}\|_{\mathrm{Lip}(X \times Y)} \leq 5c\sigma\lambda^{-1/2}$. Moreover, by Corollary 5.3 we may assume without loss of generality that $\kappa_{\psi,\varphi}$ is of the form $\kappa_{\psi,\varphi} = c_{\mathcal{Z}} \|\psi\|_{\mathrm{Lip}(X \times Y)} \|\varphi\|_{\mathrm{Lip}(X \times Y)}$, where $c_{\mathcal{Z}}$ is a constant only depending on $\mathcal{Z}$ and $(\gamma_i)$. Consequently, we obtain

$$|\mathrm{cor}_{\mathcal{Z},i}(h_{\lambda,\sigma} e_\eta^{(\sigma)}, h_{\lambda,\sigma} e_\eta^{(\sigma)})| \leq 25 c_{\mathcal{Z}} c^2 \lambda^{-1} \sigma^2 \gamma_i$$

for all $\sigma \geq 1$, $\lambda \in (0,1]$, and $\eta \in \mathbb{N}_0^d$, that is, (20) is satisfied for $K_{h_{\lambda,\sigma}} := \tilde{c}\lambda^{-1}\sigma^2$, where $\sigma \geq 1$, $\lambda \in (0,1]$, and $\tilde{c}$ is a constant independent of $\lambda$ and $\sigma$. For $n \geq 1$ and $\varepsilon > 0$ satisfying both

$$(27) \qquad \varepsilon \leq (1 + 8ea^2\sigma^2)^{-2d} |L|_{\lambda^{-1/2},1} \lambda^{-1}$$

and $\varepsilon \leq (18d \ln d)^{-2d}$, Theorem 5.11 together with (25) and (26) thus yields

$$(28) \quad \begin{aligned} \mu(\omega \in \Omega : |\mathcal{R}_{L,P}(f_{T_n(\omega),\lambda,\sigma}) &- \mathcal{R}_{L,P}(f_{P,\lambda,\sigma})| > \varepsilon) \\ &\leq \frac{2\tilde{c}(1 + 1/(8ea^2\sigma^2))^d \tilde{C}_{\lambda,\sigma,d,a}^3 |L|_{\lambda^{-1/2},1}^3 \sigma^2}{\varepsilon^3 n \lambda^4} \sum_{i=0}^{n-1} \gamma_i, \end{aligned}$$

where $\tilde{C}_{\lambda,\sigma,d,a} := (1 + \frac{1}{8ea^2\sigma^2})^{d/2} + 2\sqrt{d}e^{-a^2\sigma^2}(6a\sigma)^{(d-1)/2}|L|_{\lambda^{-1/2},1}$. Using the fact that the function $x \mapsto e^{-x^2+x}x^{(d-1)/2}$ is bounded on $[0,\infty)$ we further obtain

$$\tilde{C}_{\lambda_n,\sigma_n,d,a} \leq C_d(1 + e^{-a\sigma_n}|L|_{\lambda_n^{-1/2},1}), \tag{29}$$

where $C_d$ is a constant only depending on $d$. Let us now consider the case where Assumption S1 is fulfilled. Then we have $C_{d,a} := \sup_{n\geq 1} \tilde{C}_{\lambda_n,\sigma_n,d,a} < \infty$ and

$$\varepsilon_0 := \inf_{n\geq 1}(1 + 8ea^2\sigma_n^2)^{-2d}|L|_{\lambda_n^{-1/2},1}\lambda_n^{-1} > 0,$$

and hence (27) is satisfied for all $\varepsilon \in (0, \varepsilon_0]$. Moreover, by the remark after Assumption L we have $|L|_{\lambda^{-1/2},1} \leq c(1 + \lambda^{-1/2})$ for all $\lambda > 0$ and hence the first and third assumption of S1 together with $\sigma_n \leq \sigma_{n+1}$ imply $\lim_{n\to\infty} \lambda_n \sigma_n^d = 0$. By Lemma 5.4 we thus find $\lim_{n\to\infty} \mathcal{R}_{L,P}(f_{P,\lambda_n,\sigma_n}) = \mathcal{R}^*_{L,P}$. Consequently, (28) shows that for sufficiently large $n$ and $\varepsilon \in (0, \varepsilon_0]$ we have

$$\mu(\omega \in \Omega : |\mathcal{R}_{L,P}(f_{T_n(\omega),\lambda_n,\sigma_n}) - \mathcal{R}^*_{L,P}| > 2\varepsilon) \leq 2^{(d+1)/2}\tilde{c}C_{d,a}^3 \frac{|L|^3_{\lambda_n^{-1/2},1}\sigma_n^2}{\varepsilon^3 n \lambda_n^4} \sum_{i=0}^{n-1} \gamma_i,$$

and hence we obtain the assertion by the last condition of Assumption S1.

Let us now consider the case where Assumption S2 is fulfilled. Then it is easy to see that the second assumption of S2 implies $\lim_{n\to\infty} \sigma_n^{-4d}|L|_{\lambda_n^{-1/2},1} = 0$, which in turn yields $\sup_{n\geq 1} e^{-\sigma_n}|L|_{\lambda_n^{-1/2},1} < \infty$. From this we conclude $C_{d,a} := \sup_{n\geq 1} \tilde{C}_{\lambda_n,\sigma_n,d,a} < \infty$ by (29). Moreover, a simple consideration shows we have $\varepsilon_n := (1 + 8ea^2\sigma_n^2)^{-2d}|L|_{\lambda_n^{-1/2},1}\lambda_n^{-1} \to 0$. For a fixed $\varepsilon > 0$ we thus have $\varepsilon_n \leq \varepsilon$ for all sufficiently large $n$. Therefore we find

$$\mu(\omega \in \Omega : |\mathcal{R}_{L,P}(f_{T_n(\omega),\lambda_n,\sigma_n}) - \mathcal{R}^*_{L,P}| > 2\varepsilon)$$
$$\leq 2^{(d+1)/2}\tilde{c}C_{d,a}^3 \frac{|L|^3_{\lambda_n^{-1/2},1}\sigma_n^2}{\varepsilon_n^3 n \lambda_n^4} \sum_{i=0}^{n-1} \gamma_i \leq \tilde{C}_{d,a}\frac{\sigma_n^{2+12d}}{n\lambda_n}\sum_{i=0}^{n-1}\gamma_i$$

for all sufficiently large $n$, where $\tilde{C}_{d,a}$ is a constant only depending on $d$ and $a$. From this estimate we obtain the assertion by the last condition of Assumption S2.

Finally, let us consider the case where Assumption S3 is satisfied. Using (29) and $a \geq 1$ we then obtain for sufficiently large $n$ and $\varepsilon \in (0, \varepsilon_0]$ that

$$\mu(\omega \in \Omega : |\mathcal{R}_{L,P}(f_{T_n(\omega),\lambda_n,\sigma_n}) - \mathcal{R}^*_{L,P}| > 2\varepsilon) \leq \tilde{C}_d \frac{e^{-3\sigma_n}|L|^6_{\lambda_n^{-1/2},1}\sigma_n^2}{\varepsilon^3 n \lambda_n^4}\sum_{i=0}^{n-1}\gamma_i,$$

where $\tilde{C}_d$ is a constant only depending on $d$. $\square$

**6. Proof of Theorem 3.1.** For the proof of Theorem 3.1 we need to bound the correlation sequences for stochastic processes which are the sum of a dynamical system and an observational noise process. This is the goal of the following results. We begin with a lemma which computes the correlation of a joint process from the correlations of its components.

LEMMA 6.1. *Let $\mathcal{X} = (X_i)_{i \geq 0}$ be an $X$-valued, identically distributed process defined on $(\Omega, \mathcal{A}, \mu)$ and $\mathcal{Y} = (Y_i)_{i \geq 0}$ be a $Y$-valued, identically distributed stochastic process defined on $(\Theta, \mathcal{B}, \nu)$. Then the process $\mathcal{Z} = (Z_i)_{i \geq 0}$ defined on $(\Omega \times \Theta, \mathcal{A} \otimes \mathcal{B}, \mu \otimes \nu)$ by $Z_i := (X_i, Y_i)$ is identically distributed with $P := (\mu \otimes \nu)_{Z_0} = \mu_{X_0} \otimes \nu_{Y_0}$. Moreover, for $\psi, \varphi \in L_2(P)$ we have*

$$\mathrm{cor}_{\mathcal{Z},i}(\psi, \varphi) = \mathbb{E}_\nu \mathrm{cor}_{\mathcal{X},i}(\psi(\,\cdot\,, Y_0), \varphi(\,\cdot\,, Y_i)) + \mathbb{E}_\mu \mathbb{E}_\mu \mathrm{cor}_{\mathcal{Y},i}(\psi(X_0, \cdot), \varphi(X_0', \cdot)),$$

*where $X_0'$ is an independent copy of $X_0$.*

PROOF. The first assertion regarding $P$ is obvious. For the second assertion we fix an independent copy $\mathcal{X}' = (X_i')_{i \geq 0}$ of $\mathcal{X}$. Then an easy calculation using the fact that both $\mathcal{X}$ and $\mathcal{Y}$ are identically distributed yields

$$\begin{aligned}
&\mathrm{cor}_{\mathcal{Z},i}(\psi, \varphi) \\
&= \mathbb{E}_\mu \mathbb{E}_\nu \psi(X_0, Y_0) \varphi(X_i, Y_i) - \mathbb{E}_\mu \mathbb{E}_\nu \psi(X_0, Y_0) \cdot \mathbb{E}_\mu \mathbb{E}_\nu \varphi(X_0, Y_0) \\
&= \mathbb{E}_\mu \mathbb{E}_\nu \psi(X_0, Y_0) \varphi(X_i, Y_i) - \mathbb{E}_\mu \mathbb{E}_\mu \mathbb{E}_\nu \psi(X_0, Y_0) \varphi(X_0', Y_i) \\
&\quad + \mathbb{E}_\mu \mathbb{E}_\mu \mathbb{E}_\nu \psi(X_0, Y_0) \varphi(X_0', Y_i) - \mathbb{E}_\mu \mathbb{E}_\nu \psi(X_0, Y_0) \cdot \mathbb{E}_\mu \mathbb{E}_\nu \varphi(X_0, Y_0) \\
&= \mathbb{E}_\nu (\mathbb{E}_\mu \psi(X_0, Y_0) \varphi(X_i, Y_i) - \mathbb{E}_\mu \mathbb{E}_\mu \psi(X_0, Y_0) \varphi(X_0', Y_i)) \\
&\quad + \mathbb{E}_\mu \mathbb{E}_\mu \mathbb{E}_\nu \psi(X_0, Y_0) \varphi(X_0', Y_i) - \mathbb{E}_\mu \mathbb{E}_\mu (\mathbb{E}_\nu \psi(X_0, Y_0) \cdot \mathbb{E}_\nu \varphi(X_0', Y_0)) \\
&= \mathbb{E}_\nu (\mathbb{E}_\mu \psi(X_0, Y_0) \varphi(X_i, Y_i) - \mathbb{E}_\mu \psi(X_0, Y_0) \cdot \mathbb{E}_\mu \varphi(X_0, Y_i)) \\
&\quad + \mathbb{E}_\mu \mathbb{E}_\mu (\mathbb{E}_\nu \psi(X_0, Y_0) \varphi(X_0', Y_i) - \mathbb{E}_\nu \psi(X_0, Y_0) \cdot \mathbb{E}_\nu \varphi(X_0', Y_0)) \\
&= \mathbb{E}_\nu \mathrm{cor}_{\mathcal{X},i}(\psi(\cdot, Y_0), \varphi(\cdot, Y_i)) + \mathbb{E}_\mu \mathbb{E}_\mu \mathrm{cor}_{\mathcal{Y},i}(\psi(X_0, \cdot), \varphi(X_0', \cdot)),
\end{aligned}$$

that is, we have proved the assertion. □

The following elementary lemma establishes the Lipschitz continuity of a certain type of function which is important when considering the process that generates noisy observations of a dynamical system.

LEMMA 6.2. *Let $M \subset \mathbb{R}^d$ be a compact subset and $F : M \to M$ be a Lipschitz continuous map. For $B > 0$ and a fixed $j \in \{1, \ldots, d\}$ we write $X := M + [-B, B]^d$, $Y := \pi_j(X)$ and $Z := X \times Y$, where $\pi_j : \mathbb{R}^d \to \mathbb{R}$ denotes the $j$th coordinate projection. For $h \in \mathrm{Lip}(Z)$ and $x \in M$, $\varepsilon_0, \varepsilon_1 \in [-B, B]^d$ we define the function $\bar{h} : M \times [-B, B]^{2d} \to \mathbb{R}$ by*

(30) $$\bar{h}(x, \varepsilon_0, \varepsilon_1) := h(x + \varepsilon_0, \pi_j(F(x) + \varepsilon_1)).$$

*Then for all $x \in M$ and $\varepsilon_0, \varepsilon_1 \in [-B, B]^d$ we have*

$$\|\bar{h}(x, \cdot, \cdot)\|_{\mathrm{Lip}([-B,B]^{2d})} \leq (1 + \|F\|_{\mathrm{Lip}(M)}) \|h\|_{\mathrm{Lip}(Z)},$$

$$\|\bar{h}(\cdot, \varepsilon_0, \varepsilon_1)\|_{\mathrm{Lip}(M)} \leq \|h\|_{\mathrm{Lip}(Z)}.$$

PROOF. For $(\varepsilon_0, \varepsilon_1), (\varepsilon_0', \varepsilon_1') \in [-B, B]^d \times [-B, B]^d$ we obviously have

$$|h(x + \varepsilon_0, \pi_j(F(x) + \varepsilon_1)) - h(x + \varepsilon_0', \pi_j(F(x) + \varepsilon_1'))|$$
$$\leq \|h\|_{\mathrm{Lip}(Z)} (\|\varepsilon_0 - \varepsilon_0'\|_2^2 + |\pi_j(F(x) + \varepsilon_1) - \pi_j(F(x) + \varepsilon_1')|^2)^{1/2}$$
$$\leq \|h\|_{\mathrm{Lip}(Z)} \|(\varepsilon_0, \varepsilon_1) - (\varepsilon_0', \varepsilon_1')\|_2.$$

Analogously, for $x, x' \in M$ we have

$$|h(x + \varepsilon_0, \pi_j(F(x) + \varepsilon_1)) - h(x' + \varepsilon_0, \pi_j(F(x') + \varepsilon_1))|$$
$$\leq \|h\|_{\mathrm{Lip}(Z)} (\|x - x'\|_2^2 + |\pi_j(F(x) + \varepsilon_1) - \pi_j(F(x') + \varepsilon_1)|^2)^{1/2}$$
$$\leq \|h\|_{\mathrm{Lip}(Z)} (1 + \|F\|_{\mathrm{Lip}(M)}) \|x - x'\|_2.$$

From these estimates we easily obtain the assertions. $\square$

The next theorem bounds the correlation for functions defined by (30).

THEOREM 6.3. *Let $M \subset \mathbb{R}^d$ be compact and $F: M \to M$ be Lipschitz continuous such that the dynamical system $\mathcal{X} := (F^i)_{i \geq 0}$ has an ergodic measure $\mu$. Moreover, let $\gamma = (\gamma_i)_{i \geq 0}$ be a strictly positive sequence converging to zero such that*

(31) $$\mathrm{cor}_{\mathcal{X}}(\psi, \varphi) \in \Lambda(\gamma), \qquad \psi, \varphi \in \mathrm{Lip}(M).$$

*Furthermore, let $\mathcal{E} = (\varepsilon_i)_{i \geq 0}$ be a second-order stationary, $[-B, B]^d$-valued process on $(\Theta, \mathcal{B}, \nu)$ such that the $[-B, B]^{2d}$-valued process $\mathcal{Y} = (Y_i)_{i \geq 0}$ on $(\Theta, \mathcal{B}, \nu)$ that is defined by $Y_i(\vartheta) = (\varepsilon_i(\vartheta), \varepsilon_{i+1}(\vartheta))$, $i \geq 0$, $\vartheta \in \Theta$, satisfies*

(32) $$\mathrm{cor}_{\mathcal{Y}}(\psi, \varphi) \in \Lambda(\gamma), \qquad \psi, \varphi \in \mathrm{Lip}([-B, B]^{2d}).$$

*For a fixed $j \in \{1, \ldots, d\}$ we write $X := M + [-B, B]^d$, $Y := \pi_j(X)$, and $Z := X \times Y$. Define the process $\bar{\mathcal{Z}} = (\bar{Z}_i)_{i \geq 0}$ on $(\Omega \times \Theta, \mathcal{A} \otimes \mathcal{B}, \mu \otimes \nu)$ by $\bar{Z}_i = (F^i, \varepsilon_i, \varepsilon_{i+1})$, $i \geq 0$. Then for all $\psi, \varphi \in \mathrm{Lip}(Z)$ we have*

$$\mathrm{cor}_{\bar{\mathcal{Z}}}(\bar{\psi}, \bar{\varphi}) \in \Lambda(\gamma),$$

*where $\bar{\psi}$ and $\bar{\varphi}$ are defined by (30).*

PROOF. Let $c_\mathcal{X}$ and $c_\mathcal{Y}$ be the constants we obtain by applying (31) and (32) to Corollary 5.3. Moreover, since $\mathcal{E}$ is second-order stationary, we observe that $\mathcal{Y}$ is identically distributed. Applying Lemma 6.1 to the processes $\mathcal{X}$ and $\mathcal{Y}$ then yields

$$\begin{aligned}
&|\mathrm{cor}_{\bar{\mathcal{Z}},i}(\bar{\psi},\bar{\varphi})| \\
&\quad \leq |\mathbb{E}_\nu \mathrm{cor}_{\mathcal{X},i}(\bar{\psi}(\cdot,Y_0),\bar{\varphi}(\cdot,Y_i))| \\
&\quad\quad + |\mathbb{E}_{x\sim\mu}\mathbb{E}_{x'\sim\mu}\mathrm{cor}_{\mathcal{Y},i}(\bar{\psi}(F^0(x),\cdot),\bar{\varphi}(F^0(x'),\cdot))| \\
&\quad \leq c_\mathcal{X} \mathbb{E}_\nu \|\bar{\psi}(\cdot,\varepsilon_0,\varepsilon_1)\|_{\mathrm{Lip}(M)} \|\bar{\varphi}(\cdot,\varepsilon_i,\varepsilon_{i+1})\|_{\mathrm{Lip}(M)} \cdot \gamma_i \\
&\quad\quad + c_\mathcal{Y} \mathbb{E}_{x\sim\mu}\mathbb{E}_{x'\sim\mu} \|\bar{\psi}(x,\cdot)\|_{\mathrm{Lip}([-B,B]^{2d})} \|\bar{\varphi}(x',\cdot)\|_{\mathrm{Lip}([-B,B]^{2d})} \cdot \gamma_i \\
&\quad \leq c_\mathcal{X} \|\psi\|_{\mathrm{Lip}(Z)} \|\varphi\|_{\mathrm{Lip}(Z)} \cdot \gamma_i \\
&\quad\quad + c_\mathcal{Y}(1+\|F\|_{\mathrm{Lip}(M)}) \|\psi\|_{\mathrm{Lip}(Z)} \|\varphi\|_{\mathrm{Lip}(Z)} \cdot \gamma_i,
\end{aligned}$$

where in the last step we used Lemma 6.2. □

Note that for using the estimate of Theorem 6.3 in Lemma 5.8 it is necessary that the process $\mathcal{Y}$ be second-order stationary. Obviously, the latter is satisfied if the process $\mathcal{E}$ is stationary.

PROOF OF THEOREM 3.1. For a fixed $j \in \{1,\ldots,d\}$ we write $X := M + [-B,B]^d$ and $Y := \pi_j(X)$. Moreover, we define the $X \times Y$-valued process $\mathcal{Z} = (X_i,Y_i)_{i\geq 0}$ on $(M \times [-B,B]^{d\mathbb{N}}, \mu \otimes \nu)$ by $X_i := F^i + \pi_0 \circ S^i$ and $Y_i := \pi_j(F^{i+1} + \pi_0 \circ S^{i+1})$, and in addition, we write $P^{(j)} := (\mu \otimes \nu)_{(X_0,Y_0)}$. Let us further consider the $M \times [-B,B]^{2d}$-valued stationary process $\bar{\mathcal{Z}} := (F^i, \pi_0 \circ S^i, \pi_0 \circ S^{i+1})$ which is defined on $(M \times [-B,B]^{d\mathbb{N}}, \mu \otimes \nu)$. For $\psi,\varphi \in \mathrm{Lip}(X \times Y)$, Theorem 6.3 together with our decay of correlations assumptions then shows $|\mathrm{cor}_{\bar{\mathcal{Z}},i}(\bar{\psi},\bar{\varphi})| \leq \kappa_{\psi,\varphi}\gamma_i$ for all $i \geq 0$, where $\kappa_{\psi,\varphi} \in [0,\infty)$ is a constant independent of $i$. Moreover, our construction ensures $\mathrm{cor}_{\mathcal{Z},i}(\psi,\varphi) = \mathrm{cor}_{\bar{\mathcal{Z}},i}(\bar{\psi},\bar{\varphi})$ for all $i \geq 0$ and hence Theorem 2.4 yields

$$\mu \otimes \nu((x,\varepsilon) \in M \times [-B,B]^{d\mathbb{N}} : |\mathcal{R}_{L,P^{(j)}}(f_{T_n^{(j)}(x,\varepsilon),\lambda_n,\sigma_n}) - \mathcal{R}^*_{L,P^{(j)}}| > \varepsilon) \to 0$$

for $n \to \infty$ and all $\varepsilon > 0$. Using Assumption LD and the definition (10) we then easily obtain the assertion. □

## APPENDIX: PROOF OF THEOREM 5.1

In the following, $B_E$ denotes the closed unit ball of a Banach space $E$. Recall that a linear operator $S : E \to F$ acting between two Banach spaces $E$ and $F$ is continuous if and only if it is *bounded*, that is, $\|S\| :=$ $\sup_{x \in B_E} \|Sx\| < \infty$. Our first goal is to recall another equivalent condition which in practice is often easier to check. To this end, we need the following definition:

DEFINITION A.1. Let $E$ and $F$ be Banach spaces and $S: E \to F$ be a linear map. Then $S$ is said to have a closed graph if for all $x \in E$, $y \in F$ and all sequences $(x_n) \subset E$ satisfying $x_n \to x$ and $Sx_n \to y$ we have $Sx = y$.

Obviously, every continuous linear operator has a closed graph. The following fundamental result from functional analysis known as the *closed graph theorem* shows the converse implication.

THEOREM A.2. *Let $E$ and $F$ be Banach spaces and $S: E \to F$ be a linear map that has a closed graph. Then $S$ is continuous.*

Our next goal is to establish an analogous result for bilinear maps. To this end, we first recall the *principle of uniform boundedness*, which is also known as Banach–Steinhaus theorem.

THEOREM A.3. *Let $E$ and $F$ be Banach spaces, $A$ be a nonempty set, and $S_\alpha: E \to F$, $\alpha \in A$, be bounded linear operators such that*

$$\sup_{\alpha \in A} \|S_\alpha x\| < \infty$$

*for all $x \in E$. Then we even have $\sup_{\alpha \in A} \sup_{x \in B_E} \|S_\alpha x\| < \infty$.*

Let us now recall that a map $S: E_1 \times E_2 \to F$ between Banach spaces $E_1$, $E_2$ and $F$ is called *bilinear* if the maps $S(x_1, \cdot): E_2 \to F$ and $S(\cdot, x_2): E_1 \to F$ are linear for all $x_1 \in E_1$ and $x_2 \in E_2$. In order to state a closed graph theorem for bilinear maps we also need a notion which describes a closed graph property for bilinear maps:

DEFINITION A.4. Let $E_1$, $E_2$ and $F$ be Banach spaces and $S: E_1 \times E_2 \to F$ be a bilinear map. Then $S$ is said to have a partially closed graph if the linear maps $S(x_1, \cdot): E_2 \to F$ and $S(\cdot, x_2): E_1 \to F$ have closed graphs for all $x_1 \in E_1$ and $x_2 \in E_2$.

With these preparations we can now state and prove the announced closed graph theorem for bilinear maps:

THEOREM A.5. *Let $E_1$, $E_2$ and $F$ be Banach spaces and $S: E_1 \times E_2 \to F$ be a bilinear map that has a partially closed graph. Then there exists a constant $c \in [0, \infty)$ such that*

$$\|S(x_1, x_2)\|_F \leq c \|x_1\|_{E_1} \cdot \|x_2\|_{E_2}, \qquad x_1 \in E_1, x_2 \in E_2.$$

PROOF. By the closed graph theorem the maps $S(x_1,\cdot)\colon E_2 \to F$ and $S(\cdot,x_2)\colon E_1 \to F$ are bounded linear operators for all $x_1 \in E_1$ and $x_2 \in E_2$. In particular, the boundedness of the operators $S(\cdot,x_2)\colon E_1 \to F$ yields

$$\sup_{x_1 \in B_{E_1}} \|S(x_1,x_2)\| < \infty, \qquad x_2 \in E_2.$$

Applying the principle of uniform boundedness to the family of bounded operators $(S(x_1,\cdot))_{x_1 \in B_{E_1}}$ thus shows

$$c := \sup_{x_1 \in B_{E_1}} \sup_{x_2 \in B_{E_2}} \|S(x_1,x_2)\| < \infty.$$

Using the bilinearity of $S$ we then obtain the assertion. $\square$

With these preparations we can now present the proof of Theorem 5.1.

PROOF OF THEOREM 5.1. Obviously, $\operatorname{cor}_{\mathcal{Z}}\colon E_1 \times E_2 \to F$ is a well-defined bilinear operator. In view of Theorem A.5 it suffices to show that this operator has a partially closed graph. We begin by showing that $\operatorname{cor}_{\mathcal{Z}}(\psi,\cdot)\colon E_2 \to F$ has a closed graph for all $\psi \in E_1$. To this end let us fix some $\psi \in E_1$, $\varphi \in E_2$, a sequence $b := (b_n)_{n \geq 0} \in F$ and a sequence $(\varphi_i)_{i \geq 1} \subset E_2$ such that $\lim_{i \to \infty} \|\varphi_i - \varphi\|_{E_2} = 0$ and

$$\lim_{i \to \infty} \|\operatorname{cor}_{\mathcal{Z}}(\psi,\varphi_i) - b\|_F = 0. \tag{33}$$

Obviously, $\operatorname{cor}_{\mathcal{Z}}(\psi,\cdot)\colon E_2 \to F$ has a closed graph if $\operatorname{cor}_{\mathcal{Z}}(\psi,\varphi) = b$. To show this equality we first observe that for fixed $n \geq 0$ and $i \to \infty$ we have

$$\left|\int_Z \varphi_i \, dP - \int_Z \varphi \, dP\right| \leq \|\varphi - \varphi_i\|_{L_1(P)} \leq \|\mathrm{id}\colon E_2 \to L_2(P)\| \cdot \|\varphi - \varphi_i\|_{E_2}$$

and

$$\left|\int_\Omega \psi(Z_0) \cdot \varphi(Z_n) \, d\mu - \int_\Omega \psi(Z_0) \cdot \varphi_i(Z_n) \, d\mu\right|$$
$$\leq \|\psi\|_{L_2(P)} \cdot \|\varphi - \varphi_i\|_{L_2(P)}$$
$$\leq \|\psi\|_{L_2(P)} \cdot \|\mathrm{id}\colon E_2 \to L_2(P)\| \cdot \|\varphi - \varphi_i\|_{E_2}.$$

From this we conclude $\lim_{i \to \infty} \operatorname{cor}_{\mathcal{Z},n}(\psi,\varphi_i) = \operatorname{cor}_{\mathcal{Z},n}(\psi,\varphi)$ for the $n$th coordinate of sequences of correlations. Moreover, $F$ is continuously included in $\ell_\infty$ and hence (33) implies $\lim_{i \to \infty} \operatorname{cor}_{\mathcal{Z},n}(\psi,\varphi_i) = b_n$ for all $n \geq 0$. Combining these considerations yields $\operatorname{cor}_{\mathcal{Z},n}(\psi,\varphi) = b_n$ for all $n \geq 0$, that is, we have shown that $\operatorname{cor}_{\mathcal{Z}}(\psi,\cdot)\colon E_2 \to F$ has a closed graph. Since the fact that all $\operatorname{cor}_{\mathcal{Z}}(\cdot,\varphi)\colon E_1 \to F$ have a closed graph can be shown completely analogously, the proof is complete. $\square$

**Acknowledgment.** The authors gratefully thank V. Baladi for pointing us to the unpublished note [13] of P. Collet.

Los Alamos National Laboratory
Information Sciences CCS-3
MS B256
Los Alamos, New Mexico 87545
USA
E-mail: ingo@lanl.gov
　　　　manghel@lanl.gov